\documentclass[lettersize,journal]{IEEEtran}
\usepackage{algorithmicx}
\usepackage{algpseudocode}
\usepackage{amsmath}
\usepackage{xcolor}
\usepackage{amsmath,amsfonts}
\usepackage{algorithm}
\usepackage{array}
\usepackage[caption=false,font=normalsize,labelfont=sf,textfont=sf]{subfig}
\usepackage{textcomp}
\usepackage{stfloats}
\usepackage{url}
\usepackage{verbatim}
\usepackage{graphicx}
\usepackage{cite}
\usepackage{booktabs, makecell}
\usepackage{tabularx}
\usepackage{pifont}
\usepackage{hyperref}

\begin{document}

\title{Efficient and Expressive Public Key Authenticated Encryption with Keyword Search in Multi-user Scenarios}

\author{Jiayin Cai, Xingwen Zhao, Dexin Li, Hui Li,~\IEEEmembership{Member,~IEEE}, Kai Fan,~\IEEEmembership{Member,~IEEE}
\thanks{}
\thanks{}}

\markboth{}%
{Shell \MakeLowercase{\textit{et al.}}: A Sample Article Using IEEEtran.cls for IEEE Journals}


\maketitle

\begin{abstract}
Public key authenticated encryption with keyword search (PAEKS) represents a significant advancement of secure and searchable data sharing in public network systems, such as medical systems. It can effectively mitigate the risk of keyword guessing attacks (KGA), which is a critical issue in public key encryption with keyword search (PEKS). However, in scenarios with a large number of users, the enforced point-to-point access control necessitates that the data sender encrypt the same keyword using the public keys of multiple receivers to create indexes, while the data receiver also must generate trapdoors of size linear to senders in the system. The burden on users aiming for efficient data sharing is considerable, as the overheads increase linearly with the number of users. Furthermore, the majority of current PAEKS schemes lack expressive search functions, including conjunctions, disjunctions, or any monotone boolean formulas, which are prevalent in practical applications. To tackle the abovementioned challenges, we propose an efficient and expressive PAEKS scheme. In efficiency, one auxiliary server is integrated to assist users in generating indexes and trapdoors. Users encrypt with their respective private keys along with the public keys of the servers, facilitating secure and searchable data sharing while significantly minimizing overhead. Additionally, the LSSS is employed to implement expressive search, including monotone boolean queries. We also obfuscate the mapping relationship associated with the LSSS matrix to the keywords, thereby enhancing the privacy protection. Security analysis alongside theoretical and experimental evaluations of our scheme illustrates its practicality and efficiency in multi-user data sharing scenarios.
\end{abstract}


\section{Introduction} \label{Introduction}

In contemporary healthcare, electronic health records (EHR) have emerged as a crucial method for management in hospitals\cite{DBLP:journals/tifs/LiHHS23}\cite{DBLP:journals/tifs/YangM16}\cite{DBLP:journals/tifs/ChengM24}. The fast expansion of patient data and the rising adoption of cloud computing technologies have led numerous medical institutions to outsource the storage of EHRs to cloud servers. While it offers convenience and scalability, the role of external servers as untrusted third parties has garnered concern due to possible security risks. To provide privacy protection, data usually needs to be encrypted before outsourcing. Nevertheless, conventional encryption techniques exhibit considerable constraints in real-world implementations. For instance, when physicians in a hospital want access to patient records to formulate treatment regimens or when external medical researchers seek to query data for epidemiological studies, they must download all encrypted data to local decryption before querying and utilizing it. This incurs significant computation and communication burdens while also potentially heightening the risk of data leakage.

To perform secure keyword searches on cloud servers, Song et al.\cite{DBLP:conf/sp/SongWP00} pioneered symmetric searchable encryption technology, which has subsequently motivated further research by numerous scholars\cite{DBLP:conf/asiacrypt/AttrapadungFI06,DBLP:conf/ndss/IslamKK12,DBLP:conf/sp/NaveedPG14,DBLP:conf/ccs/Bost16,DBLP:conf/infocom/0015YWWX19,DBLP:journals/tdsc/KermanshahiLSNL21,DBLP:journals/tsc/WangSWLC22,DBLP:journals/joc/AsharovSS21,DBLP:journals/tifs/XuSWCWLJ22,DBLP:journals/tdsc/LiuXYZHLW24}. Searchable encryption enables the data sender to extract a set of keywords from the document intended for sharing and subsequently upload the encrypted keywords as an index alongside the ciphertext document to the cloud server for storage. To initiate a query for a document, the receiver encrypts the keyword using his/her key to generate a query trapdoor and transmit it to the cloud server, which then executes matching algorithms and returns the results. During this time, the cloud is not allowed to learn any information regarding the keywords. Searchable encryption algorithms are categorized into two types based on the cryptographic systems employed for generating indexes and trapdoors: symmetric searchable encryption (SSE) and asymmetric searchable encryption (ASE), the latter also referred to as public key encryption with keyword search (PEKS). PEKS was first proposed by Boneh et al.\cite{DBLP:conf/eurocrypt/BonehCOP04}. In contrast to SSE, it eliminates the necessity for users to exchange keys beforehand, making it especially appropriate for public network settings. It has been extensively utilized in applications such as electronic health records, email systems, and the Internet of Things\cite{DBLP:journals/tdsc/ZhangXWZW21,DBLP:journals/tifs/YangM16,DBLP:journals/tmc/LuL22,DBLP:conf/uss/Meng0TML24,DBLP:journals/iandc/CachetADRHF23}, etc.

Although PEKS has a wide range of applications, Byun et al.\cite{DBLP:conf/sdmw/ByunRPL06} noted that it is typically susceptible to keyword guessing attacks (KGA), in which the adversary selects a keyword for encryption and performs an equality test algorithm upon receiving a trapdoor. If the results match, the keyword is revealed; otherwise, the adversary proceeds to select another keyword for encryption and testing. One of the reasons this attack is successful is that the entropy of the keyword space in the searchable encryption scheme is low, allowing the attacker to easily traverse it. KGA is classified into external KGA and internal KGA; the former is initiated by an external attacker who obtains the trapdoors illegally, whereas the latter is directly initiated by a malicious test server. So far, a variety of methods have been proposed to resist KGA, including designated servers\cite{DBLP:journals/jss/RheePSL10}, fuzzy keyword search\cite{DBLP:journals/tc/XuJWW13}, dual servers\cite{DBLP:journals/tifs/ChenMYGW16}, registered keyword search\cite{DBLP:conf/europki/TangC09}, etc. But one primary reason KGA is effective is that any individual can utilize the receiver's public key to encrypt the keyword. Huang and Li\cite{DBLP:journals/isci/HuangL17} came up with the idea of public key authenticated encryption with keyword search (PAEKS), which requires the sender's private key to be input when encrypting, making it harder for attackers to directly create ciphertext index and launch KGA. Currently, PAEKS has emerged as one of the most promising searchable encryption schemes capable of withstanding KGA.

\begin{figure*}
	\centering
	\includegraphics[width=1\linewidth]{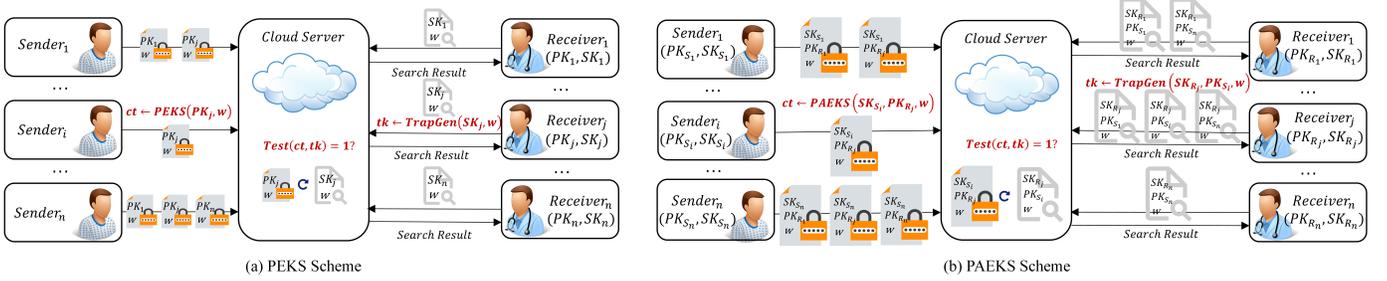}
	\caption{PEKS and PAEKS.\label{fig:1}}
\end{figure*}

While PAEKS schemes are KGA-resistant, their limitations regarding
client-side computational efficiency are evident in multi-user scenarios, especially the EHR application system. Specifically, before a sender (such as a patient) uploads and shares data with multiple receivers (such as researchers or doctors in research institutions), he/she needs to encrypt the keywords using his/her private key and multiple receivers' public keys, respectively, thereby generating multiple indexes. This feature is derived from the foundational architecture of PEKS, leading to a notable rise in encryption computational overhead with an increasing number of receivers, as illustrated in Figure \ref{fig:1} (a). Additionally, in PAEKS, when a receiver intends to query EHRs that include a specific keyword (for example, "heart disease"), he/she must first acquire the public keys of all senders associated with that keyword and then utilize his/her own private key and multiple senders' public keys to generate the trapdoors, the quantity of which produced exhibits a linear correlation with the number of senders, as illustrated in Figure \ref{fig:1} (b). Therefore, it is essential to design a scheme to keep KGA-secure while optimizing computational and communication efficiency in large-scale user scenarios.

To tackle the above issues, Li et al.\cite{DBLP:journals/tifs/LiHHS23} introduced public key authenticated encryption with ciphertext update and keyword search (PAUKS), which achieves constant trapdoor overhead in multi-user environments. This is accomplished by incorporating a trusted administrator within the system to re-encrypt the sender's ciphertext, thereby transforming it into a format that is independent of the original sender. While the design of Li et al.'s scheme is effective, as pointed out by Cheng et al.\cite{DBLP:journals/tifs/ChengM24}, it still results in a system bottleneck due to the participation of a trusted administrator, who possesses the authority to access all keywords queried by each receiver and must be entirely trustworthy. Thus, Cheng et al. enhanced their scheme by innovatively using a sender server as well as a receiver server, functioning as agents for multiple senders and receivers, respectively. When generating ciphertext indexes and trapdoors, the sender and receiver are required to utilize only the public key of the other side's server rather than each user's, which maintains the constant ciphertext and query trapdoor overhead effectively.

Cheng et al.\cite{DBLP:journals/tifs/ChengM24} have effectively addressed the computation and communication overhead on the user side; however, we observed that most current PEKS, including Cheng's scheme, support only single keyword search functions, which is inadequate for EHR systems that require more expressive search, which is defined as the keywords queried that can be expressed as conjunctions, disjunctions, or any monotone boolean formulas \cite{DBLP:conf/uss/Meng0TML24}. For instance, in a statistical survey on the relationship between disease and age, researchers might issue a query "(disease: hypertension $\mathbf{OR}$ diabetes)" $\mathbf{AND}$ (age: 40 $\mathbf{OR}$ 50). To date, schemes supporting expressive queries have been put forward \cite{DBLP:journals/tdsc/CuiWDWL18,DBLP:conf/eurocrypt/KatzSW08,DBLP:conf/ccs/LaiZDLC13,DBLP:conf/isw/LvHZF14,DBLP:conf/provsec/MengZNLHS17,DBLP:journals/access/ShenLL20,DBLP:journals/istr/TsengFL22,DBLP:journals/tdsc/HeGWWLY20,rao2023boolean}, but either of them have complex designs that lead to inefficiency or do not consider KGA attacks. Meng et al.\cite{DBLP:conf/uss/Meng0TML24} introduced the first PAEKS scheme that can implement expressive search. However, in large-scale user scenarios, the above-mentioned user computing and communication overhead problems still cannot be solved. In addition, the adopted partial hiding policy in some schemes results in the exposure of keyword names on the cloud server, thereby leaking the sensitive information in EHR. For example, if terms like "delivery outcome" and "postpartum heart rate" are present, it becomes straightforward to deduce the patient's personal information, potentially compromising the privacy.

\subsection{Contributions}
We propose a PAEKS scheme that is both efficient and expressive based on the aforementioned observations. Inspired by the recent work of Cheng et al.\cite{DBLP:journals/tifs/ChengM24}, we introduce the incorporation of a semi-trusted auxiliary server within a three-party architecture comprising senders, receivers, and a cloud server. The auxiliary server is responsible for re-encrypting the ciphertext index and query trapdoor prior to submission to the untrusted cloud server and can be modeled as an internal search management server within the hospital EHR system, which is restricted from colluding with the external cloud storage server. Users encrypt the keyword with their private key and the public keys of two servers, thereby minimizing computation and communication overhead regardless of the number of users. Furthermore, to facilitate expressive boolean queries, we utilize the concept of KP-ABE, as implemented by Meng et al.’s state of the art\cite{DBLP:conf/uss/Meng0TML24}, to convert the receiver's monotone boolean query into an LSSS matrix, with each row representing a distinct keyword. Our scheme treats user keyword names as highly sensitive information; thus, we provide optional keyword privacy protection. Specifically, our contributions are as follows:
\begin{itemize}
	\item \textbf{EE-PAEKS.} We propose an Efficient and Expressive Public key Authenticated Encryption scheme with Keyword Search (EE-PAEKS) applicable to multi-user scenarios. We formally define the system model and security model in Section \ref{System and Security Models}, and give the specific construction in Section \ref{Our Scheme}.
	\item \textbf{Efficiency.} The introduction of a single auxiliary server addresses the administrator bottleneck of Li et al.\cite{DBLP:journals/tifs/LiHHS23}, while maintaining equivalent user computation and communication efficiency as in \cite{DBLP:journals/tifs/ChengM24}, with reduced server overhead. To a certain extent, we consider the internal auxiliary server and the external cloud server to the destination users of senders and receivers, and it is required that the two servers not collude with each other. Furthermore, similar to \cite{DBLP:journals/tifs/LiHHS23} and \cite{DBLP:journals/tifs/ChengM24}, our scheme allows the cloud server to create an inverted index offline to enhance search efficiency, which is discussed in detail in Section \ref{Further Discussion}.
	\item \textbf{Expressiveness.} To support expressive search functions, we introduce the LSSS matrix converted from KP-ABE to effectively enrich the receiver's search capabilities. The sender uses multiple keywords (resp. attribute in KP-ABE) to generate encrypted indexes, and the receivers convert the boolean query, even threshold query, into the LSSS keyword policy (resp. attribute policy) matrix, in which each keyword is represented by $\{name, value\}$. Only files that satisfy the queries will be correctly located and returned to the receiver.
	\item \textbf{Security and performance evaluation.} We present the threat and security model of EE-PAEKS in Section \ref{System and Security Models} and provide a formal security proof in Section \ref{Our Scheme}. In addition, we conceal the plaintext keyword names that are frequently revealed in other schemes. Given the significance of privacy in systems like EHR, we emphasize and provide the protection of keyword names in our scheme. Specifically, we encrypt the mapping values in the keyword policy matrix in trapdoors, which is optional.  In Section \ref{Performance}, we present the theoretical and experimental analysis of the scheme. We implement our system within the Charm framework and conduct a comparative analysis with the state-of-the-art. Experiments show that our scheme has comparable performance in terms of user-side overhead for flexible search in multi-user scenarios, making it potential in applications.
\end{itemize}

\subsection{Related Works}

\textit{PEKS of different public key systems.} In practice, PEKS requires a trusted center to generate and manage the user's public key in PKI, leading to certificate management problems. To solve this problem, Abdalla et al.\cite{DBLP:conf/crypto/AbdallaBCKKLMNPS05} introduced identity-based encryption with keyword search (IBEKS), in which the user's identity information can be used as a public key. However, a fully trusted Key Generation Center (KGC) remains necessary to generate the corresponding user private key, thereby resulting in the key-escrow problem. Additionally, some IBEKS cannot resist KGA. Therefore, inspired by Huang and Li's PAEKS\cite{DBLP:journals/isci/HuangL17}, Li et al.\cite{DBLP:journals/isci/LiHSYS19} proposed identity-based authenticated encryption with keyword search (IBAEKS). In order to solve the key-escrow problem, Peng et al.\cite{yanguo2014certificateless} first proposed certificateless encryption (CLEKS) in 2014. Similarly, to solve the KGA faced, the concept of CLAEKS was first expanded by He et al.\cite{DBLP:journals/tii/HeMZKL18}, and the first security model was defined. However, CLAEKS is limited in its use in public networks because it requires a KGC to distribute part-private keys to system users through a secure channel. Therefore, Lu et al.\cite{DBLP:journals/tsc/LuLZ21} extended the concept of PEKS to the certificate-based encryption (CBC) setting and defined the concept of certificate-based encryption with keyword search (CBEKS), which subsequently inspired works such as \cite{DBLP:conf/provsec/YangLHZAH22,DBLP:conf/provsec/LiuLYSTH19,DBLP:journals/jsa/ShiralyPNE22,DBLP:journals/iotj/ChengM23,DBLP:journals/tii/LuLW21,DBLP:journals/istr/ShiralyEP24}.

\textit{PEKS with expressive search.} Most PEKS with expressive search \cite{DBLP:journals/tdsc/CuiWDWL18,DBLP:conf/ccs/LaiZDLC13,DBLP:conf/isw/LvHZF14,DBLP:conf/provsec/MengZNLHS17,DBLP:journals/access/ShenLL20,DBLP:journals/istr/TsengFL22} are based on anonymous key policy attribute-based encryption (A-KP-ABE). Key policy attribute-based encryption (KP-ABE) \cite{DBLP:conf/ccs/GoyalPSW06} is a public key cryptography that supports fine-grained access control. The sender generates ciphertext by encrypting an attribute set while the receiver uses the key bound to the access policy to decrypt the ciphertext. A successful decryption requires that the attribute set in the ciphertext satisfies the access policy. Linear secret sharing technology\cite{DBLP:phd/il/Beimel96} is usually used in KP-ABE to support expressive access policies. To enhance privacy protection, A-KP-ABE is used to hide the attribute names in the ciphertext. In the PEKS scheme with expressive search, the keyword set and the trapdoor generated by the access policy correspond to the attribute set and the key bound to the access policy in A-KP-ABE, respectively. However, existing schemes still have limitations\cite{DBLP:conf/ccs/LaiZDLC13,DBLP:conf/isw/LvHZF14,DBLP:conf/eurocrypt/KatzSW08,DBLP:journals/tdsc/CuiWDWL18,DBLP:conf/uss/Meng0TML24}. For example, the schemes based on composite order groups in \cite{DBLP:conf/ccs/LaiZDLC13} and \cite{DBLP:conf/isw/LvHZF14} have high computational overhead, and the scheme based on inner product encryption in \cite{DBLP:conf/eurocrypt/KatzSW08} shows super-polynomial growth in the size of ciphertext and trapdoor, which limits its practical application. In recent years, Cui et al.\cite{DBLP:journals/tdsc/CuiWDWL18} and Meng et al.\cite{DBLP:conf/uss/Meng0TML24} have proposed more efficient and secure solutions, but as far as we know, the problem of protecting keyword privacy and reducing user computation and communication overhead while supporting expressive search in large-scale user scenarios has not been effectively solved yet.

\textit{Large-scale user-efficient PEKS.} In large-scale user scenarios, the computational efficiency issues of PEKS and PAEKS are particularly prominent. To address this challenge, Venkata et al.\cite{DBLP:journals/istr/ChenamA23} proposed a scheme that enables the sender to generate encrypted keyword indexes for multiple users once and supports joint keyword queries, thereby significantly improving computational efficiency. Liu et al.\cite{DBLP:conf/acisp/LiuHYSTH21} designed a broadcast PAEKS scheme that allows all receivers in the broadcast list to search for ciphertexts simultaneously, further optimizing the efficiency of multi-user scenarios. Li et al.\cite{DBLP:journals/tifs/LiHHS23} proposed a PAEKS scheme that supports ciphertext updates and fast keyword search and improved user efficiency by introducing a fully trusted administrator using proxy re-encryption technology. Subsequently, Cheng et al.\cite{DBLP:journals/tifs/ChengM24} improved the scheme by adopting a dual-server architecture to solve the bottleneck problem of the fully trusted administrator. However, there is currently no high-user-efficiency PAEKS scheme that supports expressive queries for EHR scenarios, which inspired our work in this paper.

\section{Preliminaries} \label{Preliminaries}
\subsection{Bilinear Map}
Let $\mathbb{G}_1,\mathbb{G}_2$, and $\mathbb{G}_T$ be three multiplicative groups of prime order $p$. A bilinear pairing is a mapping $e:\mathbb{G}_1 \times \mathbb{G}_2\rightarrow \mathbb{G}_T$ satisfies the following properties:
\begin{itemize}
	\item $\mathbf{Bilinearity}$: $e(u^a,v^b)=e(u,v)^{ab}$ for any $u\in \mathbb{G}_1,v\in \mathbb{G}_2$ and $a,b\in \mathbb{Z}_p$.
	\item $\mathbf{Nondegeneracy}$: There exist $u\in \mathbb{G}_1,v\in \mathbb{G}_2$ such that $e(u,v)\neq I$ where $I$ is the identity element of $\mathbb{G}_T$.
	\item $\mathbf{Computability}$: For all $u\in \mathbb{G}_1,v\in \mathbb{G}_2$, $e$ can be efficiently computed.
\end{itemize}

\subsection{Modified Decision Linear Problem}
The modified Decision Linear (mDLIN) problem \cite{boneh2004short}\cite{DBLP:conf/icics/BaoDZ03} is described as follows:

Given a generator $g\in \mathbb{G}_1$ , elements $g^a,g^b,g^{ac},g^{d/b}, \mathcal{Z} \in \mathbb{G}_1$, decide whether $\mathcal{Z}$ is equal to $g^{c+d}$ or a randomly selected $g^z$, where $a,b,c,d$ are randomly chosen from $\mathbb{Z}_p$.

\textit{ Definition 1 (mDLIN Assumption):}
The mDLIN assumption holds if the above mDLIN problem is intractable for any probabilistic polynomial-time (PPT) adversary.

\subsection{LSSS}
\textit{ Definition 2 (Access Structure):} Suppose the set of attributes is $Q=\left\{Q_1,Q_2,…,Q_n\right\}$,
$\Gamma\subseteq 2^{(|Q|)}$ is monotone if $X\in \Gamma$ and $X\subseteq Y$, then $Y\in \Gamma$ for $\forall X,Y$. The nonempty subset $\Gamma$  of $Q$ is an access structure if the attribute sets in $\Gamma$ are considered as
authorized sets, and otherwise, they are treated as unauthorized.

\textit{ Definition 3 (LSSS):}
Given a monotonic access structure $(M,\rho )$, where $M_{l\times n}$ is the shared generator matrix and $\rho$ maps each row of the matrix to an attribute. An \textit{LSSS} scheme includes the following two algorithms:

\textit{1) Secret sharing algorithm}: First, randomly select the secret value to be shared as $s \in Z_p$, and then choose $n-1$ random values ${\{r_2,...r_n\}}$ from $Z_p$, forming an $n$-dimensional column vector $\vec{r}={\{s,r_2,...r_n\}}$. Then calculate $\lambda_i=M_i\cdot \vec{r}$, where $M_i$ is the row vector formed by the $i$th row of $M$ and $\lambda_i$ represents the $i$th shared share of the secret value $s$.

\textit{2) Secret Reconstructing algorithm}: Enter the secret sharing value set $\{\lambda _i\}_{i=\{1,...l\}}$ and the attribute set $S$, and symbol the access policy as $\mathbb{A}$. Suppose $S$ is the authorization set, that is, $S \in  \mathbb{A}$, let $I=\{\rho (i)\in \mathbb{A} \subseteq [1,...l]\}$. If $\{\lambda _i\}_{i=\{1,...l\}}$ is a valid set for $s$, then there is a set of constants $\{\omega _i|i\in I\}$ such that $\sum_{i\in I}\omega _iM_i=(1,0,...,0)$ is satisfied. These constants $\{\omega _i\}_{i\in I}$ can be found in polynomial time with the size
of the share-generating matrix $M$, and for unauthorized sets, no such constants $\{\omega _i\}_{i\in I}$ exist. And then the secret value can be recovered as $s=\sum_{i\in I}\omega _i \lambda _i$.

It is worth noting that we use keyword sets (resp. keyword policy) to correspond to the attribute sets (resp. attribute policy) in related concepts in KP-ABE. The related concepts in the two encryption systems are essentially not distinguished. In addition, there are many algorithms for constructing the policy matrix in the linear secret sharing scheme (LSSS), among which the most common ones are the Lewko-Waters algorithm\cite{DBLP:conf/eurocrypt/LewkoW11a} and the Liu-Cao-Wong algorithm\cite{liu2010efficient}. We employ the Liu-Cao-Wong algorithm for matrix construction for the following reasons: 1) Compared with the Lewko-Waters algorithm, it is more general and supports thresholds for non-leaf nodes in the access tree. 2) The first column elements of the generated matrix are all 1, which provides the foundation for the construction of our security analysis; see \ref{Our Scheme} for details.

\section{System and Security Models} \label{System and Security Models}

\subsection{EE-PAEKS Framework}

Our EE-PAEKS is composed of the following algorithms:

\begin{itemize}
	\item $\mathbf{Setup(1^\lambda)}\to \mathbb{PP}$.
	Given the security parameter $1^\lambda$, the algorithm returns the public parameter $\mathbb{PP}$. 
	\item $\mathbf{KeyGen}_{C}( \mathbb{PP}) \to ( pk_{C}, sk_{C})$. Given the public parameter $\mathbb{PP}$, this algorithm returns the public and secret keys $(pk_{C},sk_{C})$ for cloud server.
	\item $\mathbf{KeyGen}_{A}( \mathbb{PP}) \to ( pk_{A}, sk_{A})$. Given the public parameter $\mathbb{PP}$, this algorithm returns the public and secret keys $(pk_{A},sk_{A})$ for auxiliary server.
	\item $\mathbf{KeyGen}_{S}( \mathbb{PP}) \to ( pk_{S}, sk_{S})$. Given the public parameter $\mathbb{PP}$, this algorithm returns sender's public and secret keys $( pk_{S}, sk_{S})$.
	\item $\mathbf{KeyGen}_{R}( \mathbb{PP}) \to ( pk_{R}, sk_{R})$. Given the public parameter $\mathbb{PP}$, this algorithm returns receiver's public and secret keys $( pk_{R}, sk_{R})$.
	\item $\mathbf{Enc}(\mathbb{PP},sk_{S},pk_{C},pk_{A},\mathbb{W})\to ct$. Given the public parameter $\mathbb{PP}$, a sender's secret key $sk_{S}$, cloud server's public key $pk_{C}$, auxiliary server's public key $pk_{A}$ and a set of keywords $\mathbb{W}$, this algorithm returns a ciphertext $ct$.
	\item$\mathbf{EncTrans}(sk_{A},ct)\to ct'$. Given the auxiliary server's secret key $sk_{A}$ and sender's ciphertext $ct$, this algorithm returns a transformed ciphertext $ct'$.
	\item $\mathbf{Trap}( \mathbb{PP}, sk_{R}, pk_{C},pk_{A}, \mathbb{P}) \rightarrow td$. Given the public parameter $\mathbb{PP}$, a receiver's secret key $sk_{R}$, cloud server's public key $pk_{C}$, auxiliary server's public key $pk_{A}$ and a keyword policy $\mathbb{P}$, this algorithm returns a trapdoor $td$.
	\item$\mathbf{TrapTrans}(sk_{A},td)\to td'$. Given the auxiliary server's secret key $sk_{A}$ and receiver's trapdoor $td$, this algorithm returns a transformed trapdoor $td'$.
	\item $\mathbf{Search}(sk_{C},ct',td')\to0/1$. Given the cloud server's secret key $sk_{C}$, transformed ciphertext $ct'$ and transformed trapdoor $td'$, this algorithm output 1 if the search is succesful; Otherwise, it returns 0.
\end{itemize}

\begin{figure}[htbp]
	\centering
	\includegraphics[width=0.45\textwidth]{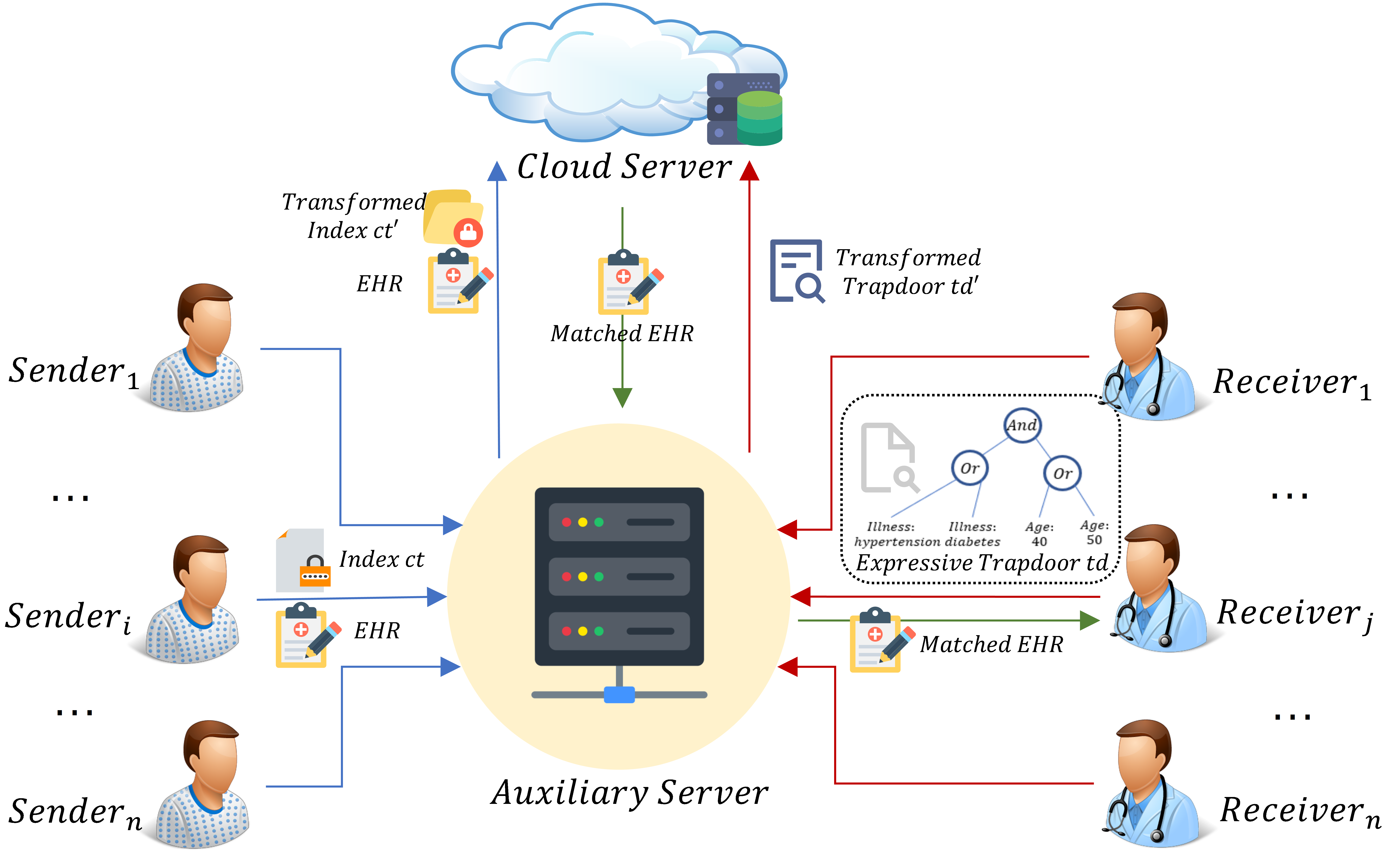}
	\caption{System Model.\label{fig:2}}
\end{figure}

\subsection{System Model}
As shown in Figure 2, our system model consists of four main entities: multiple senders, multiple receivers, an auxiliary server, and a cloud server. The role and capabilities of each entity in the system are as follows:

\begin{itemize}
	\item \textbf{Senders:} Each sender executes $\mathbf{KeyGen_S(\cdot)}$ to generate his/her own public-private key pair, and runs $\mathbf{Enc}(\cdot)$ to encrypt the keyword set $\mathbb{W}$ and generate the ciphertext index $ct$. Then, he/she sends the encrypted EHR record along with $ct$ to the auxiliary server for further encryption.
	\item \textbf{Receivers:} Each receiver executes $\mathbf{KeyGen_R(\cdot)}$ to generate his/her own public-private key pair, and runs $\mathbf{Trap}(\cdot)$ to encrypt the keyword policy $\mathbb{P}$ converted by boolean query and generate trapdoor $td$. Then, he/she sends $td$ to the auxiliary server for further encryption.
	\item \textbf{Auxiliary Server (AS):} The auxiliary server first executes $\mathbf{KeyGen_A(\cdot)}$ to generate its own public-private key pair. It transforms ciphertext $ct$ from a sender to $ct'$ by running $\mathbf{EncTrans(\cdot)}$ and sends it along with the EHR record to cloud server for storage. When a receiver initiates a search request, it runs $\mathbf{TrapTrans(\cdot)}$ to transform trapdoor $td$ to $td'$ and sends it to the cloud server.
	\item \textbf{Cloud Server (CS):} The cloud server first executes $\mathbf{KeyGen_C(\cdot)}$ to generate its own public-private key pair. When receiving a search request from the AS, it executes $\mathbf{Search}(\cdot)$ to determine whether the transformed ciphertext meets the keyword policy of the transformed trapdoor, and returns the search result.
\end{itemize}

\begin{table}[!t]
	\caption{Notations\label{tab1}}
	\centering
	\begin{tabular}{|c|c|}
		\hline
		Notation & Description\\
		\hline
		$1^\lambda$ & Security parameter\\
		\hline
		$\mathbb{PP}$ & Public parameter\\
		\hline
		$p$ & A large prime number\\
		\hline
		$g$ & Generator of group $\mathbb{G}$\\
		\hline
		$\mathbb{G},\mathbb{G}_T$ & Multiplicative cyclic groups with order $p$\\
		\hline
		$H$ & A hash function $H:\{0,1\}^* \to \mathbb{G}$ \\
		\hline
		$(pk_S,sk_S)$ & Data sender's key pairs \\
		\hline
		$(pk_R,sk_R)$ & Data receiver's key pairs \\
		\hline
		$(pk_C,sk_C)$ & Cloud server's key pairs \\
		\hline
		$(pk_A,sk_A)$ & Auxiliary server's key pairs \\
		\hline
		$ct$ & Keyword ciphertext \\
		\hline
		$td$ & Search trapdoor \\
		\hline
		$ct'$ & Transformed keyword ciphertext  \\
		\hline
		$td'$ & Transformed search trapdoor \\
		\hline
		$\mathbb{W}$ & \makecell*[l]{A keyword set $\mathbb{W}=\{\omega_i\}_{i\in[m]}=\{n_i,v_i\}_{i \in[m]}$,\\ where $n_i$ is keyword name and $v_i$ is keyword value}\\
		\hline
		$\mathbb{P}$ & \makecell*[l]{A keyword policy $\mathbb{P}=\{M,\pi,\{\pi(i)\}_{i\in[l]}\}$, where\\$M$ is the policy matrix, $\pi$ is a mapping of each row \\of the matrix to a keyword and remind that $\{\pi(i)\}_{i \in [l]}$ \\  $=\{n_{\pi(i)},v_{\pi(i)}\}_{i\in[l]}$}\\
		\hline
	\end{tabular}
\end{table}

\subsection{Threat Model}
In our scheme, both the cloud server and the auxiliary server are considered to be honest but curious, they will correctly execute the algorithm in the system, but may try to obtain the privacy concealed in ciphertext and trapdoors. Collusion is not allowed between these two servers. This is realistic in the EHR scenario because the cloud server provides a third-party outsourced storage service, while the auxiliary server is modeled as a local server within the hospital system. In addition, similar to \cite{DBLP:journals/tifs/ChengM24}, we allow the auxiliary server to collude with malicious users other than the challenge sender/receiver, but do not allow the cloud server to collude with any malicious users.
\subsection{Security Model}
In this section, similar to \cite{DBLP:journals/tifs/ChengM24}, we extend the ciphertext indistinguishability (CI) and trapdoor indistinguishability (TI) in \cite{DBLP:journals/isci/HuangL17}, where the former is defined as the ciphertext does not reveal any information about the keyword set (including keyword names), the latter is defined as the trapdoor does not reveal any information about the keyword policy (also including keyword names) in our system. We formally define the security model of the system by considering the following games between the challenger $\mathcal{C}$ and the adversary $\mathcal{A}$.

\textit{1) Ciphertext Indistinguishability Against Auxiliary Server (CI-AS):}

\textbf{Initialization:} Given a security parameter $1^\lambda$, $\mathcal{C}$ generates the public parameter $\mathbb{PP}$. It then runs $\mathbf{KeyGen}_{C}$,  $\mathbf{KeyGen}_{S}$ and $\mathbf{KeyGen}_{R}$ to generate the key pairs $(pk_C,sk_C),(pk_S,sk_S),(pk_R,sk_R)$ respectively, and sends $(\mathbb{PP},pk_C,pk_S,pk_R)$ to $\mathcal{A}$. Next, $\mathcal{A}$ plays the role of the auxiliary server by running $\mathbf{KeyGen}_{A}( \mathbb{PP}) \to ( pk_{A}, sk_{A})$ and sending $pk_{A}$ to $\mathcal{C}$.

\textbf{Query phase \uppercase\expandafter{\romannumeral 1:} } $\mathcal{A}$ is allowed to adaptively issue queries to the following oracles for polynomially many times:
\begin{itemize}
	\item \textbf{Ciphertext Oracle} $\mathcal{O}_{ct}:$ Given a keyword set $\mathbb{W}$, $\mathcal{C}$ runs the algorithm $\mathbf{Enc}(\mathbb{PP},sk_{S},pk_{C},pk_{A},\mathbb{W})\to ct$ and returns $ct$ to $\mathcal{A}$.
	\item \textbf{Trapdoor Oracle} $\mathcal{O}_{td}:$ Given a keyword policy $\mathbb{P}$, $\mathcal{C}$ runs the algorithm $\mathbf{Trap}( \mathbb{PP}, sk_{R}, pk_{C},pk_{A}, \mathbb{P}) \rightarrow td$ and returns $td$ to $\mathcal{A}$. 
\end{itemize}

\textbf{Challenge:} $\mathcal{A}$ selects two keyword sets of equal size, $\mathbb{W}_0^*=\{n_{i0},v_{i0}\}_{i\in[m]}$ and $\mathbb{W}_1^*=\{n_{i1},v_{i1}\}_{i\in[m]}$, and submits them to $\mathcal{C}$. Next, $\mathcal{C}$ flips a random coin $b \in \{0,1\}$ and runs $\mathbf{Enc}(\mathbb{PP},sk_{S},pk_{C},pk_{A},\mathbb{W}_b^*)\to ct_b^*$, then returns $ct_b^*$ to $\mathcal{A}$.

\textbf{Query phase \uppercase\expandafter{\romannumeral 2:} } $\mathcal{A}$ continues to issue queries, as in Query Phase \uppercase\expandafter{\romannumeral 1}.

\textbf{Guess:} $\mathcal{A}$ outputs the guess $b' \in \{0,1\}$ and wins the game if $b'=b$. 

In the CI-AS game, neither of $\mathbb{W}_0^*$ and $\mathbb{W}_1^*$ is allowed to be queried in $\mathcal{O}_{ct}$ or satisfy any policy that has been queried in $\mathcal{O}_{td}$. We define the advantage of $\mathcal{A}$ to win the CI-AS game as

$$\mathsf{Adv}_{\mathcal{A}}^{CI-AS}(1^\lambda)=\left | Pr[b'=b]-\frac{1}{2}\right |.$$

\textit{ Definition 4:} If $\mathsf{Adv}_{\mathcal{A}}^{CI-AS}(1^\lambda)$ is negligible in $1^\lambda$ for any PPT adversary, then the proposed scheme is CI-AS secure.

\textit{2) Ciphertext Indistinguishability Against Cloud Server (CI-CS):}

\textbf{Initialization:} Given a security parameter $1^\lambda$, $\mathcal{C}$ generates the public parameter $\mathbb{PP}$. It then runs $\mathbf{KeyGen}_{A}$,  $\mathbf{KeyGen}_{S}$ and $\mathbf{KeyGen}_{R}$ to generate the key pairs $(pk_A,sk_A),(pk_S,sk_S),(pk_R,sk_R)$ respectively, and sends $(\mathbb{PP},pk_A,pk_S,pk_R)$ to $\mathcal{A}$. Next, $\mathcal{A}$ plays the role of the cloud server by running $\mathbf{KeyGen}_{C}( \mathbb{PP}) \to ( pk_{C}, sk_{C})$ and sending $pk_{C}$ to $\mathcal{C}$.

\textbf{Query phase \uppercase\expandafter{\romannumeral 1:} } In addition to the two queries in the CI-AS game, $\mathcal{A}$ is allowed to adaptively issue two other queries to the following oracles for polynomially many times:
\begin{itemize}
	\item \textbf{Trans-Ciphertext Oracle} $\mathcal{O}_{Tran_{ct}}:$ Given a ciphertext $ct$, $\mathcal{C}$ runs the algorithm $\mathbf{EncTrans}(sk_{A},ct)\to ct'$ and returns $ct'$ to $\mathcal{A}$.
	\item \textbf{Trans-Trapdoor Oracle} $\mathcal{O}_{Tran_{td}}:$ Given a trapdoor $td$, $\mathcal{C}$ runs the algorithm $\mathbf{TrapTrans}(sk_{A},td)\to td'$ and returns $td'$ to $\mathcal{A}$. 
\end{itemize}

\textbf{Challenge:} $\mathcal{A}$ selects two keyword sets of equal size, $\mathbb{W}_0^*=\{n_{i0},v_{i0}\}_{i\in[m]}$ and $\mathbb{W}_1^*=\{n_{i1},v_{i1}\}_{i\in[m]}$, and submits them to $\mathcal{C}$ as challenge keyword sets. Next, $\mathcal{C}$ flips a random coin $b \in \{0,1\}$ and runs $\mathbf{Enc}(\mathbb{PP},sk_{S},pk_{C},pk_{A},\mathbb{W}_b^*)\to ct_b^*$, then returns $ct_b^*$ to $\mathcal{A}$.

\textbf{Query phase \uppercase\expandafter{\romannumeral 2:} } $\mathcal{A}$ continues to issue queries, as in Query Phase \uppercase\expandafter{\romannumeral 1}.

\textbf{Guess:} $\mathcal{A}$ outputs the guess $b' \in \{0,1\}$ and wins the game if $b'=b$. 

We require that, in any Query phase of the CI-CS game, $\mathcal{A}$ should not have queried any $(n_{i0},v_{i0})$ or $(n_{i1},v_{i1})$ in $\mathcal{O}_{ct}$ or $\mathcal{O}_{td}$.

We define the advantage of $\mathcal{A}$ to win the CI-CS game as:

$$\mathsf{Adv}_{\mathcal{A}}^{CI-CS}(1^\lambda)=\left | Pr[b'=b]-\frac{1}{2}\right |.$$

\textit{ Definition 5:} If $\mathsf{Adv}_{\mathcal{A}}^{CI-CS}(1^\lambda)$ is negligible in $1^\lambda$ for any PPT adversary, then the proposed scheme is CI-CS secure.

\textit{3) Trapdoor Indistinguishability Against Auxiliary Server (TI-AS):} The TI-AS game is defined almost the same as the CI-AS game except for the challenge phase.

\textbf{Challenge:} $\mathcal{A}$ selects two keyword policies of equal size, $\mathbb{P}_0^*=\{M_0^*,\pi_0^*,\{n_{\pi_0^*(i)},v_{\pi_0^*(i)}\}_{i\in[l]}$ and $\mathbb{P}_1^*=\{M_1^*,\pi_1^*,\{n_{\pi_1^*(i)},v_{\pi_1^*(i)}\}_{i\in[l]}$, and submits them to $\mathcal{C}$ as the challenge keyword policies. Next, $\mathcal{C}$ flips a random coin $b \in \{0,1\}$ and runs $\mathbf{Trap}( \mathbb{PP}, sk_{R}, pk_{C},pk_{A}, \mathbb{P}_b^*) \rightarrow td_b^*$, then returns $td_b^*$ to $\mathcal{A}$.

\textbf{Query phase \uppercase\expandafter{\romannumeral 2:} } $\mathcal{A}$ continues to issue queries as in Query Phase \uppercase\expandafter{\romannumeral 1}.

In the TI-AS game, neither of $\mathbb{P}_0^*$ and $\mathbb{P}_1^*$ is allowed to be queried in $\mathcal{O}_{td}$ or be satisfied by any keyword set that has been queried in $\mathcal{O}_{ct}$. We define the advantage of $\mathcal{A}$ to win the TI-AS game as

$$\mathsf{Adv}_{\mathcal{A}}^{TI-AS}(1^\lambda)=\left | Pr[b'=b]-\frac{1}{2}\right |.$$

\textit{ Definition 6:} If $\mathsf{Adv}_{\mathcal{A}}^{TI-AS}(1^\lambda)$ is negligible in $1^\lambda$ for any PPT adversary, then the proposed scheme is TI-AS secure.

\textit{4) Trapdoor Indistinguishability Against Cloud Server (TI-CS):} The TI-CS game is defined almost the same as the CI-CS game except for the challenge phase.

\textbf{Challenge:} $\mathcal{A}$ selects two keyword policies of equal size, $\mathbb{P}_0^*=\{M_0^*,\pi_0^*,\{n_{\pi_0^*(i)},v_{\pi_0^*(i)}\}_{i\in[l]}$ and $\mathbb{P}_1^*=\{M_1^*,\pi_1^*,\{n_{\pi_1^*(i)},v_{\pi_1^*(i)}\}_{i\in[l]}$, and submits them to $\mathcal{C}$ as the challenge keyword policies. Next, $\mathcal{C}$ flips a random coin $b \in \{0,1\}$ and runs $\mathbf{Trap}( \mathbb{PP}, sk_{R}, pk_{C},pk_{A}, \mathbb{P}_b^*) \rightarrow td_b^*$, then returns $td_b^*$ to $\mathcal{A}$.

We require that, in any Query phase of the TI-CS game, $\mathcal{A}$ should not have queried any $(n_{\pi_0^*(i)},v_{\pi_0^*(i)})$ or $(n_{\pi_1^*(i)},v_{\pi_1^*(i)})$ in $\mathcal{O}_{ct}$ or $\mathcal{O}_{td}$.

We define the advantage of $\mathcal{A}$ to win the TI-CS game as

$$\mathsf{Adv}_{\mathcal{A}}^{TI-CS}(1^\lambda)=\left | Pr[b'=b]-\frac{1}{2}\right |.$$

\textit{ Definition 7:} If $\mathsf{Adv}_{\mathcal{A}}^{TI-CS}(1^\lambda)$ is negligible in $1^\lambda$ for any PPT adversary, then the proposed scheme is TI-CS secure.

\section{Construction of EE-PAEKS scheme} \label{Our Scheme}

In this section, we describe the construction details of our scheme.
\begin{itemize}
	\item $\mathbf{Setup(1^\lambda)}\to \mathbb{PP}$.
	Given the security parameter $1^\lambda$, choose a hash function $H:\{0,1\}^* \to \mathbb{G}$ and output the public parameter $\mathbb{PP}=\{p,g, \mathbb{G}, \mathbb{G}_T,e,H\}$, where $p$ is a large prime, $\mathbb{G}$ and $\mathbb{G}_T$ are groups with order $p$ and $e:\mathbb{G}\times \mathbb{G} \to \mathbb{G} _T$ is a bilinear map. 
	
	\item $\mathbf{KeyGen}_{C}( \mathbb{PP}) \to ( pk_{C}, sk_{C})$. Given the public parameter $\mathbb{PP}$, pick $c\in \mathbb{Z}_p^*$ randomly and output the public key $pk_{C}=g^{c}$ and  secret key $sk_{C}=c$ for the cloud server.
	
	\item $\mathbf{KeyGen}_{A}( \mathbb{PP}) \to ( pk_{A}, sk_{A})$. Given the public parameter $\mathbb{PP}$, pick $a\in \mathbb{Z}_p^*$ randomly and output the public key $pk_{A}=g^{a}$ and  secret key $sk_{A}=a$ for the auxiliary server.
	
	\item $\mathbf{KeyGen}_{S}( \mathbb{PP}) \to ( pk_{S}, sk_{S})$. Given the public parameter $\mathbb{PP}$, pick $s\in \mathbb{Z}_p^*$ randomly and output the public key $pk_{S}=g^{s}$ and secret key $sk_{S}=s$ for a data sender.
	
	\item $\mathbf{KeyGen}_{R}( \mathbb{PP}) \to ( pk_{R}, sk_{R})$. Given the public parameter $\mathbb{PP}$, pick $r\in \mathbb{Z}_p^*$ randomly and output the public key $pk_{R}=g^{r}$ and  secret key $sk_{R}=r$ for a data receiver.
	
	\item$\mathbf{Enc}(\mathbb{PP},sk_{S},pk_{C},pk_{A},\mathbb{W}=\{\omega_i\}_{i\in[m]}=\{n_i,v_i\}_{i \in[m]})\to ct$. Given the public parameter $\mathbb{PP}$, a sender's secret key $sk_{S}$, cloud server's public key $pk_{C}$ and auxiliary server's public key $pk_{A}$, the algorithm randomly selects $x_1,x_2,x_3\in \mathbb{Z}_p^*$ and outputs the ciphertext $ct=(\{ct_{1,i}\}_{i\in [m]},ct_2,ct_3,ct_4)$, where
	$$ct=\begin{cases}ct_{1,i}=H(\omega_i)^{sk_{S}x_3}g^{x_1+x_2},\\ct_2=pk_{C}^{x_1},\\ct_3=pk_{A}^{x_2},\\ct_4=g^{sk_{S}x_3}.\end{cases}$$
	
	\item $\mathbf{EncTrans}(sk_{A},ct)\to ct'$. Given auxiliary server's secret key $sk_{A}$ and sender's ciphertext $ct$, the algorithm randomly selects $\hat{x}\in \mathbb{Z}_p^*$ and outputs the transformed ciphertext $ct'=(\{ct_{1,i}'\}_{i\in [m]},ct_2',ct_3',ct_4')$ as follows:
	$$ct'=\begin{cases}ct_{1,i}'=ct_{1,i}^{\hat{x}{sk_A}},\\ct_2'=ct_2^{\hat{x}sk_{A}},\\ct_3'=ct_3^{\hat{x}},\\ct_4'=ct_4^{\hat{x}sk_A}.\end{cases}$$

	\item $\mathbf{Trap}( \mathbb{PP}, sk_{R},pk_{C},pk_{A},\mathbb{P}=\{M,\pi,\{\pi(i)\}_{i\in[l]}\}) \rightarrow td$. Remind that $\{\pi(i)\}_{i \in [l]}=\{n_{\pi(i)},v_{\pi(i)}\}_{i\in[l]}$. Given the public parameter $\mathbb{PP}$, a receiver's secret key $sk_{R}$, cloud server's public key $pk_{C}$ and auxiliary server's public key $pk_{A}$, the algorithm randomly selects $y_1,y_2,y_3\in \mathbb{Z}_p^*$ and $\textbf{v}\in \mathbb{Z}_p^{n-1}$ and outputs the trapdoor $td=(\mathbb{P}=\{M,\pi,\{td_{12,i}\}_{i\in[l]}\},\{td_{11,i}\}_{i\in [l]},td_2,td_3,td_4)$ as follows:
	$$td=\begin{cases}td_{11,i}=g^{M_{i}(y_{1}+y_{2}\|\textbf{v})^{\mathrm{T}}}\cdot H(\pi(i))^{sk_Ry_{3}},\\
	td_{12,i}=g^{y_{1}+y_{2}}\cdot H(\pi(i))^{sk_Ry_{3}},\\td_{2}={pk_C}^{y_{1}},\\td_{3}={pk_A}^{y_{2}},\\td_{4}=g^{sk_Ry_{3}}.\end{cases}$$
	Note that the mapped value $\{\pi(i)\}_{i\in[l]}$ is replaced by $\{td_{12,i}\}_{i\in[l]}$ to fully hide the keyword names and values.
	
	\item $\mathbf{TrapTrans}(sk_{A},td)\to td'$. Given auxiliary server's secret key $sk_{A}$ and receiver's trapdoor $td$, the algorithm randomly selects $\hat{y}\in \mathbb{Z}_p^*$ and outputs the transformed trapdoor $td'=(\mathbb{P}'=\{M,\pi,\{td_{12,i}'\}_{i\in[l]}\},\{td_{11,i}'\}_{i\in [l]},td_2',td_3',td_4')$ as follows:
	$$\begin{aligned}&td^{\prime}=\begin{cases}td_{11,i}^{\prime}=td_{11,i}^{\hat{y}sk_{A}},\\td_{12,i}'=td_{12,i}^{\hat{y}sk_A},\\td_{2}^{\prime}=td_{2}^{\hat{y}sk_A},\\td_{3}^{\prime}=td_{3}^{\hat{y}},\\td_{4}^{\prime}=td_{4}^{\hat{y}sk_A}.\end{cases}\end{aligned}$$
	The mapped value $\{td_{12,i}\}_{i\in[l]}$ is re-encrypted by auxiliary server into $\{td_{12,i}'\}_{i\in[l]}$.
	
	\item $\mathbf{Search}(sk_{C},ct',td')\to0/1$. Given cloud server's secret key $sk_{C}$, sender's transformed ciphertext $ct'$ and receiver's transformed trapdoor $td'$, the algorithm executes the following steps:

	(1) Test whether there exists a subset $I'$ in $ct'$ that satisfies the policy $\mathbb{P}'$ in $td'$.
	First, calculate $\delta_i=e(\frac{ct_{1,i}^{\prime_{sk_C}}}{ct_{2}^{\prime}ct_{3}^{\prime_{sk_C}}},td_{4}^{\prime})$ for each $\{ct_{1,i}^{\prime}\}_{i\in[m]}$ in $ct'$. Similarly, calculate $\mu_{j}=e(\frac{td_{12,j}^{\prime sk_C}}{td_{2}^{\prime}td_{3}^{\prime sk_C}},ct_{4}^{\prime})$ for each $\{td_{12,j}^{\prime}\}_{j\in[l]}$ in $td'$. If there exists $\delta_i=\mu_{j}$, then the corresponding matrix index row number is recorded; Otherwise, return 0.

	(2) Find a constant set $\{\gamma_k\}_{k\in I^{\prime}}$ s.t. $\sum_{k\in I^{\prime}}\gamma_kM_k=(1,0,...,0)$ and calculate:
	$$e(\frac{\prod_{k\in I'}(td_{11,k}')^{\gamma_ksk_C}}{td_2'td_3'^{sk_C}},ct_4')=e(\prod_{k\in I'}(\frac{ct_{1,k}'^{sk_C}}{ct_2'ct_3'^{sk_C}})^{\gamma_k},td_4')$$
	If the equation holds, it returns 1. Otherwise, the cloud server continues to look for another subset and repeats the check step. If the above equation does not hold for all subsets, it returns 0.
\end{itemize}

\subsection{Correctness}
If a constant set $\{\gamma_k\}_{k\in I^{\prime}}$ s.t. $\sum_{k\in I^{\prime}}\gamma_kM_k=(1,0,...,0)$ is found, the cloud server calculates as follows:

\begin{math}
\begin{aligned}
&e(\frac{\prod_{k\in I^{\prime}}(td_{11,k}^{\prime})^{\gamma_{k}sk_C}}{td_{2}^{\prime}td_{3}^{\prime sk_C}},ct_{4}^{\prime})\\
&=e(\frac{\prod_{k\in I^{\prime}}(g^{\hat{y}M_{k}(y_{1}+y_{2}\|\textbf{v})^{\mathrm{T}}sk_A}\cdot H(\pi(k))^{\hat{y}y_{3}sk_Ask_R})^{\gamma_{k}sk_C}}{g^{\hat{y}y_{1}sk_Csk_A}g^{\hat{y}y_{2}sk_Ask_C}},\\
&\hspace{0.8cm}g^{\hat{x}x_{3}sk_Ssk_A})\\
&=e(\frac{g^{\hat{y}sk_Ask_C\sum_{k\in I'}\gamma_{k} M_{k}(y_{1}+y_{2}\|\textbf{v})^{\mathrm{T}}}}{g^{\hat{y}sk_Csk_A(y_{1}+y_2)}} \\
&\hspace{0.8cm}\cdot \prod_{k\in I^{\prime}}H(\pi(k))^{\hat{y}y_{3}sk_Ask_Rsk_C\gamma_{k}},g^{\hat{x}x_{3}sk_Ssk_A})\\
&=\prod_{k\in I^{\prime}}e(H(\pi(k))^{\hat{y}y_{3}sk_Ask_Rsk_C\gamma_{k}},g^{\hat{x}x_{3}sk_Ssk_A})\\
&=\prod_{k\in I^{\prime}}e(H(\pi(k)),g)^{\hat{x}x_{3}\hat{y}y_{3}sk_Csk_A^2sk_Ssk_R\gamma_{k}}\\
\end{aligned}
\end{math}

\begin{math}
\begin{aligned}
&e(\prod_{k\in I^{\prime}}(\frac{c{t_{1,k}}^{{\prime}sk_C}}{c{t_{2}}^{\prime}c{t_{3}}^{{\prime}sk_C}})^{\gamma_{k}},t{d_{4}}^{\prime})\\
&=e(\prod_{k\in I^{\prime}}(\frac{H(\omega_{k})^{\hat{x}x_{3}sk_Ssk_Csk_A}g^{\hat{x}(x_{1}+x_{2})sk_Csk_A}}{g^{\hat{x}x_{1}sk_Csk_A}g^{\hat{x}x_{2}sk_Csk_A}})^{\gamma_{k}},\\
&\hspace{0.8cm}g^{\hat{y}y_3sk_Rsk_A})\\
&=e(\prod_{k\in I^{\prime}}(H(\omega_{k})^{\hat{x}x_{3}sk_Ssk_Csk_A})^{\gamma_{k}},g^{\hat{y}y_3sk_Rsk_A})\\
&=e(\prod_{k\in I^{\prime}}(H(\omega_{k})^{\hat{x}x_{3}sk_Ssk_Csk_A\gamma_{k}}),g^{\hat{y}y_3sk_Rsk_A})\\
&=\prod_{k\in I^{\prime}}e(H(\omega_{k})^{\hat{x}x_{3}sk_Ssk_Csk_A\gamma_{k}},g^{\hat{y}y_3sk_Rsk_A})\\
&=\prod_{k\in I^{\prime}}e(H(\omega_{k}),g)^{\gamma_{k}\hat{x}x_{3}\hat{y}y_3sk_Ssk_Rsk_Csk_A^2}\\
\end{aligned}
\end{math}

If the keyword set $\mathbb{W}$ satisfies the keyword policy $\mathbb{P}$, the above two formulas are equivalent.
\subsection{Security Proof}

\textit{Theorem 1:} The proposed scheme is CI-AS secure assuming the hardness of the mDLIN problem.

\textit{Proof:} If there exists a PPT adversary $\mathcal{A}$ who can break the CI-AS security game with a non-negligible advantage $\varepsilon$, then we can construct a PPT algorithm $\mathcal{B}$ to solve the mDLIN problem. Given an mDLIN instance $\{g,g^a,g^b,g^{ac},g^{d/b},\mathcal{Z}\}$, $\mathcal{B}$'s task is to distinguish whether $\mathcal{Z}$ is $g^{c+d}$ or a randomly selected $g^z$. Let $\theta \in \{0,1\}$ such that $\theta=0$ if $\mathcal{Z}=g^{c+d}$ and $\theta=1$ otherwise. Then $\mathcal{B}$ simulates $\mathcal{A}$'s CI-AS security game as follows.

\textbf{Initialization:} $\mathcal{B}$ runs $\mathbf{Setup(1^\lambda)}\to \mathbb{PP}$, randomly chooses $r \in \mathbb{Z}_p^*$, sets $pk_R=g^r,sk_R=r,pk_C=g^a,pk_S=g^b,$ (which implies $sk_C=a,sk_S=b$). Then it sends $(\mathbb{PP},pk_C,pk_S,pk_R)$ to $\mathcal{A}$, who then submits $pk_A$ to $\mathcal{B}$.

	\textbf{Query phase \uppercase\expandafter{\romannumeral 1:} } For simplicity, we assume that: (1)$\mathcal{A}$ issues at most $\mathcal{Q}_{H}, \mathcal{Q}_{ct}, \mathcal{Q}_{td}$ queries to the hash oracle $\mathcal{O}_{H}$, the ciphertext oracle $\mathcal{O}_{ct}$ and the trapdoor oracle $\mathcal{O}_{td}$, respectively; (2)$\mathcal{A}$ does not repeat a query to an oracle; (3)$\mathcal{A}$ should issue a keyword $w$ to $\mathcal{O}_{H}$ before issuing it to $\mathcal{O}_{ct}$ and $\mathcal{O}_{td}$. 
	
	The oracles are simulated by $\mathcal{B}$ as follows: 
\begin{itemize}
	\item \textbf{Hash Oracle} $\mathcal{O}_{H}:$ $\mathcal{B}$ maintains a list $L_H$ with entries of the form $(w_i,h_i,t_i,\delta_i)$, which is initially empty. Upon receiving a keyword string $w_i$, if it has been submitted to $\mathcal{O}_{H}$, $\mathcal{B}$ retrieves the corresponding $h_i$; otherwise, $\mathcal{B}$ randomly selects $t_i \in \mathbb{Z}_p$ and a bit $\delta_i \in \{0,1\}$ such that $Pr[\delta_i=0]=\delta$, sets $h_i=g^{d/b}\cdot g^{t_i}$ if $\delta_i=0$ and $h_i=g^{t_i}$ if $\delta_i=1$, then adds $(w_i,h_i,t_i,\delta_i)$ to $L_H$ and responds with $h_i$.
	\item \textbf{Ciphertext Oracle} $\mathcal{O}_{ct}:$ Given a keyword set $\mathbb{W}=\{\omega_j\}_{j\in[m]}=\{n_j,v_j\}_{j \in[m]}$, $\mathcal{B}$ retrieves entry $(w_j,h_j,t_j,\delta_j)$ corresponding to each keyword in $\mathbb{W}$ from $L_H$. If there is an entry with $\delta_j=0$, $\mathcal{B}$ outputs a random guess for $\theta$ and aborts (event $\textsf{abort}_1$). Otherwise, $\mathcal{B}$ randomly picks $x_1,x_2,x_3 \in \mathbb{Z}_p^*$, sets 
	
	$$ct_{1,j}=(g^b)^{x_3t_j}\cdot g^{x_1+x_2}, ct_2=(g^a)^{x_1},$$	
	 $$ct_3=(pk_A)^{x_2}, ct_4=(g^b)^{x_3},$$
	and outputs $ct=(\{ct_{1,j}\}_{j\in [m]},ct_2,ct_3,ct_4)$.
	 
	\item \textbf{Trapdoor Oracle} $\mathcal{O}_{td}:$ Given a keyword policy $\mathbb{P}=\{M,\pi,\{\pi(j)\}_{j\in[l]}\}$, $\mathcal{B}$ retrieves entry $(w_j,h_j,t_j,\delta_j)$ corresponding to each keyword in $\mathbb{P}$ from $L_H$. If there is an entry with $\delta_j=0$, $\mathcal{B}$ outputs a random guess for $\theta$ and aborts (event $\textsf{abort}_1$). Otherwise, $\mathcal{B}$ randomly selects $y_1,y_2,y_3\in \mathbb{Z}_p^*$ and $\textbf{v}\in \mathbb{Z}_p^{n-1}$, sets 
	
	$$td_{11,j}=g^{M_{j}(y_{1}+y_{2}\|\textbf{v})^{\mathrm{T}}}\cdot (g^r)^{y_3t_j}, td_{12,j}=(g^r)^{y_3t_j}\cdot g^{y_1+y_2},$$	
	$$td_2=(g^a)^{y_1}, td_3=(pk_A)^{y_2}, td_4=(g^r)^{y_3},$$
	and outputs $td=(\mathbb{P}=\{M,\pi,\{td_{12,j}\}_{j\in[l]}\},\{td_{11,j}\}_{j\in [l]},td_2,td_3,td_4)$.
\end{itemize}

\textbf{Challenge:} $\mathcal{A}$ selects two keyword sets of equal size, $\mathbb{W}_0^*=\{\omega_{0i}^*\}_{i\in[n]}$ and $\mathbb{W}_1^*=\{\omega_{1i}^*\}_{i\in[n]}$ with the restriction that neither of them is allowed to be queried in $\mathcal{O}_{ct}$ or satisfy any policy that has been queried in $\mathcal{O}_{td}$. Then $\mathcal{A}$ submits them to $\mathcal{B}$. Next, $\mathcal{B}$ retrieves entry $(w_{ki},h_{ki},t_{ki},\delta_{ki})$ corresponding to each keyword in $\{\mathbb{W}_k^*\}_{k\in\{0,1\}}$ from $L_H$, where $i\in[n]$. We set $\{\Delta_k^*=null\}_{k\in\{0,1\}}$. If each $\delta_{ki}$ in the entry in $L_H$ corresponding to each $w_{ki}$ in $\mathbb{W}_k^*$ is 0, we set $\Delta_k^*=0$; Otherwise, we set $\Delta_k^*=1$. Next, $\mathcal{B}$ works as follows.

(1) If $\Delta_0^*=\Delta_1^*=1$ or there exists $\Delta_k^*=null$, where $k\in\{0,1\}$, $\mathcal{B}$ aborts (event $\textsf{abort}_2$);

(2) Otherwise, there's a bit $\hat{\omega}$ such that $\Delta_{\hat{\omega}}^*=0$ with $h_{\hat{\omega}i}^*=g^{d/b}\cdot g^{t_{\hat{\omega}i}^*}$. Then $\mathcal{B}$ generates challenge ciphertext for $\mathbb{W}_{\hat{\omega}}^*$. It randomly selects $x_2^*,x_3^* \in \mathbb{Z}_p^*$, implicitly sets $x_1^*=cx_3^*$, sets

	$$ct_{1,i}^*=\mathcal{Z}^{x_3^*}\cdot(g^b)^{x_3^*t_{\hat{\omega}i}^*}\cdot g^{x_2^*}, ct_2^*=(g^{ac})^{x_3^*},$$	
	$$ct_3^*=(pk_A)^{x_2^*}, ct_4^*=(g^b)^{x_3^*},$$

and outputs $ct^*=(\{ct_{1,i}^*\}_{i\in [n]},ct_2^*,ct_3^*,ct_4^*)$. 

If $\mathcal{Z}=g^{c+d}$, then $ct^*$ is a well-formed ciphertext; else $\mathcal{Z}=g^z$ where $z$ is randomly chosen from $\mathbb{Z}_p^*$.

\textbf{Query phase \uppercase\expandafter{\romannumeral 2:} } $\mathcal{A}$ continues to issue queries with the same restriction in the challenge phase.

\textbf{Guess:} $\mathcal{A}$ returns the guess $\hat{\omega}' \in \{0,1\}$. If $\hat{\omega}'=\hat{\omega}$, $\mathcal{B}$ sets $\theta' =0$; otherwise, it set $\theta' =1$. 

Assume that the average number of keywords in each $\mathcal{O}_{ct}$ and $\mathcal{O}_{td}$ is $m$, the keywords number of challenge ciphertext is $n$. Denote by $\textsf{abort}$ the event that $\mathcal{B}$ aborts during the game, then we have $\Pr[\overline{\textsf{abort}}]=\Pr[\overline{\textsf{abort}_1}] \cdot \Pr[\overline{\textsf{abort}_2}]=(1-\delta)^{m(\mathcal{Q}_{ct}+ \mathcal{Q}_{td})}\cdot2\delta^n(1-\delta)^n$. When $\delta=\frac{n}{2n+m(\mathcal{Q}_{ct}+\mathcal{Q}_{td})} $, the probability $\Pr[\overline{\textsf{abort}}]$ takes the maximum value, which is approximately equal to $2(\frac{1}{2+\mathcal{Q}_{ct}+\mathcal{Q}_{td}} )^n$ when $m$ and $n$ are close and thus non-negligible. Therefore, the probability that $\mathcal{B}$ succeeds in guessing the bit $\theta$ and solves the mDLIN problem is 
\begin{equation*}
\begin{aligned}
\Pr[\theta'=\theta]&=\Pr[\theta'=\theta\wedge {\textsf{abort}}]+\Pr[\theta'=\theta\wedge \overline{\textsf{abort}}] \\
&= \frac{1}{2}(1-\Pr[\overline{\textsf{abort}}])+(\frac{1}{2}+\epsilon )\cdot \Pr[\overline{\textsf{abort}}]  \\
&= \frac{1}{2}+ \epsilon \cdot \Pr[\overline{\textsf{abort}}].\\
\end{aligned}
\end{equation*}

If $\epsilon$ is non-negligible, so is $ |\Pr[\theta'=\theta]-\frac{1}{2}|$.

\textit{Theorem 2:} The proposed scheme is CI-CS secure assuming the hardness of the mDLIN problem.

\textit{Proof:} The proof is similar to that of Theorem 1.

\textbf{Initialization:} $\mathcal{B}$ runs $\mathbf{Setup(1^\lambda)}\to \mathbb{PP}$, randomly chooses $r \in \mathbb{Z}_p^*$, sets $pk_R=g^r,sk_R=r,pk_A=g^a,pk_S=g^b,$ (which implies $sk_A=a,sk_S=b$). Then it sends $(\mathbb{PP},pk_A,pk_S,pk_R)$ to $\mathcal{A}$, who then submits $pk_C$ to $\mathcal{B}$.

\textbf{Query phase \uppercase\expandafter{\romannumeral 1:} } $\mathcal{A}$ submits polynomial queries to $\mathcal{O}_{H},\mathcal{O}_{ct},\mathcal{O}_{td},\mathcal{O}_{Tran_{ct}},\mathcal{O}_{Tran_{td}}.$ $\mathcal{B}$ responds $\mathcal{O}_{H}$ as in Theorem 1 and responds other oracles as follows. 

\begin{itemize}

	\item \textbf{Ciphertext Oracle} $\mathcal{O}_{ct}:$ Given a keyword set $\mathbb{W}=\{\omega_j\}_{j\in[m]}=\{n_j,v_j\}_{j \in[m]}$, $\mathcal{B}$ retrieves entry $(w_j,h_j,t_j,\delta_j)$ corresponding to each keyword in $\mathbb{W}$ from $L_H$. If there is an entry with $\delta_j=0$, $\mathcal{B}$ outputs a random guess for $\theta$ and aborts (event $\textsf{abort}_1$). Otherwise, $\mathcal{B}$ randomly picks $x_1,x_2,x_3 \in \mathbb{Z}_p^*$, sets 
	
	$$ct_{1,j}=(g^b)^{x_3t_j}\cdot g^{x_1+x_2}, ct_2=(pk_C)^{x_1},$$	
	$$ct_3=(g^a)^{x_2}, ct_4=(g^b)^{x_3},$$
	and $ct=(\{ct_{1,j}\}_{j\in [m]},ct_2,ct_3,ct_4)$, adds $(ct,\{t_j\}_{j\in[m]},x_1,x_2,x_3)$ in an initially empty list $L_{ct}$ and returns $ct$ to $\mathcal{A}$.
	
	\item \textbf{Transform Ciphertext Oracle} $\mathcal{O}_{Tran_{ct}}:$ Given a ciphertext $ct=(\{ct_{1,j}\}_{j\in [m]},ct_2,ct_3,ct_4)$, $\mathcal{B}$ retrieves entry $(ct,\{t_j\}_{j\in[m]},x_1,x_2,x_3)$ from $L_{ct}$. $\mathcal{B}$ randomly picks $x' \in \mathbb{Z}_p^*$ and implicitly sets $x'=\hat{x}a$, then sets
	
	$$ct_{1,j}'=ct_{1,j}^{x'}, ct_2'=ct_2^{x'},ct_3'=(g^{x_2})^{x'}, ct_4'=ct_4^{x'},$$
	and outputs $ct'=(\{ct_{1,j}'\}_{j\in [m]},ct_2',ct_3',ct_4')$.
	
	\item \textbf{Trapdoor Oracle} $\mathcal{O}_{td}:$ Given a keyword policy $\mathbb{P}=\{M,\pi,\{\pi(j)\}_{j\in[l]}\}$, $\mathcal{B}$ retrieves entry $(w_j,h_j,t_j,\delta_j)$ corresponding to each keyword in $\mathbb{P}$ from $L_H$. If there is an entry with $\delta_j=0$, $\mathcal{B}$ outputs a random guess for $\theta$ and aborts (event $\textsf{abort}_1$). Otherwise, $\mathcal{B}$ randomly selects $y_1,y_2,y_3\in \mathbb{Z}_p^*$ and $\textbf{v}\in \mathbb{Z}_p^{n-1}$, sets 
	
	$$td_{11,j}=g^{M_{j}(y_{1}+y_{2}\|\textbf{v})^{\mathrm{T}}}\cdot (g^r)^{y_3t_j}, td_{12,j}=(g^r)^{y_3t_j}\cdot g^{y_1+y_2},$$	
	$$td_2=pk_C^{y_1}, td_3=(g^a)^{y_2}, td_4=(g^r)^{y_3},$$
	
	and $td=(\mathbb{P}=\{M,\pi,\{td_{12,j}\}_{j\in[l]}\},\{td_{11,j}\}_{j\in [l]},td_2,td_3,td_4)$, adds $(td,\{t_j\}_{j\in[m]},y_1,y_2,y_3,\textbf{v})$  in an initially empty list $L_{td}$ and returns $td$ to $\mathcal{A}$.
	
	\item \textbf{Transform Trapdoor Oracle} $\mathcal{O}_{Tran_{ts}}:$ Given a trapdoor $td=(\mathbb{P}=\{M,\pi,\{td_{12,j}\}_{j\in[l]}\},\{td_{11,j}\}_{j\in [l]},td_2,td_3,td_4)$, $\mathcal{B}$ retrieves entry $(td,\{t_j\}_{j\in[m]},y_1,y_2,y_3,\textbf{v})$  from $L_{td}$. $\mathcal{B}$ randomly picks $y' \in \mathbb{Z}_p^*$ and implicitly sets $y'=\hat{y}a$, then sets
	
$$td_{11,j}'=td_{11,j}^{y'}, td_{12,j}'=td_{12,j}^{y'},$$	
$$td_2'=td_2^{y'}, td_3'=(g^{y_2})^{y'}, td_4'=td_4^{y'},$$

	and outputs $td'=(\mathbb{P}'=\{M,\pi,\{td_{12,j}'\}_{j\in[l]}\},\{td_{11,j}'\}_{j\in [l]},td_2',td_3',td_4')$.

\end{itemize}

\textbf{Challenge:} $\mathcal{A}$ selects two keyword sets of equal size, $\mathbb{W}_0^*=\{\omega_{0i}^*\}_{i\in[n]}$ and $\mathbb{W}_1^*=\{\omega_{1i}^*\}_{i\in[n]}$ with the restriction that $\mathcal{A}$ should not have queried any $\omega_{0i}^*$ or $\omega_{1i}^*$ in $\mathcal{O}_{ct}$ or $\mathcal{O}_{td}$. Next, $\mathcal{B}$ retrieves entry $(w_{ki},h_{ki},t_{ki},\delta_{ki})$ corresponding to each keyword in $\{\mathbb{W}_k^*\}_{k\in\{0,1\}}$ from $L_H$, where $i\in[n]$. We set $\{\Delta_k^*=null\}_{k\in\{0,1\}}$. If each $\delta_{ki}$ in the entry in $L_H$ corresponding to each $w_{ki}$ in $\mathbb{W}_k^*$ is 0, we set $\Delta_k^*=0$; Otherwise, we set $\Delta_k^*=1$. Next, $\mathcal{B}$ works as follows.

(1) If $\Delta_0^*=\Delta_1^*=1$ or there exists $\Delta_k^*=null$, where $k\in\{0,1\}$, $\mathcal{B}$ aborts (event $\textsf{abort}_2$);

(2) Otherwise, there's a bit $\hat{\omega}$ such that $\Delta_{\hat{\omega}}^*=0$ with $h_{\hat{\omega}i}^*=g^{d/b}\cdot g^{t_{\hat{\omega}i}^*}$. Then $\mathcal{B}$ generates challenge ciphertext for $\mathbb{W}_{\hat{\omega}}^*$. It randomly selects $x_1^*,x_3^* \in \mathbb{Z}_p^*$, implicitly sets $x_2^*=cx_3^*$, sets

$$ct_{1,i}^*=\mathcal{Z}^{x_3^*}\cdot(g^b)^{x_3^*t_{\hat{\omega}i}^*}\cdot g^{x_1^*}, ct_2^*=(pk_C)^{x_1^*},$$	
$$ct_3^*=(g^{ac})^{x_3^*}, ct_4^*=(g^b)^{x_3^*},$$

and outputs $ct^*=(\{ct_{1,i}^*\}_{i\in [n]},ct_2^*,ct_3^*,ct_4^*)$. 

If $\mathcal{Z}=g^{c+d}$, then $ct^*$ is a well-formed ciphertext; else $\mathcal{Z}=g^z$ where $z$ is randomly chosen from $\mathbb{Z}_p^*$.

\textbf{Query phase \uppercase\expandafter{\romannumeral 2:} } $\mathcal{A}$ continues to issue queries with the same restriction in the challenge phase.

\textbf{Guess:} $\mathcal{A}$ returns the guess $\hat{\omega}' \in \{0,1\}$. If $\hat{\omega}'=\hat{\omega}$, $\mathcal{B}$ sets $\theta' =0$; otherwise, it set $\theta' =1$. 

Similar to Theorem 1, the probability that $\mathcal{B}$ succeeds in guessing the bit $\theta$ and solves the mDLIN problem is $ |\Pr[\theta'=\theta]-\frac{1}{2}|$, which is non-negligible.

\textit{Theorem 3:} The proposed scheme is TI-AS secure assuming the hardness of the mDLIN problem.

\textit{Proof:} The proof is similar to that of Theorem 1.

\textbf{Initialization:} $\mathcal{B}$ runs $\mathbf{Setup(1^\lambda)}\to \mathbb{PP}$, randomly chooses $s \in \mathbb{Z}_p^*$, sets $pk_S=g^s,sk_S=s,pk_C=g^a,pk_R=g^b,$ (which implies $sk_C=a,sk_R=b$). Then it sends $(\mathbb{PP},pk_C,pk_S,pk_R)$ to $\mathcal{A}$, who then submits $pk_A$ to $\mathcal{B}$.

\textbf{Query phase \uppercase\expandafter{\romannumeral 1:} } $\mathcal{A}$ submits polynomial queries to $\mathcal{O}_{H},\mathcal{O}_{ct},\mathcal{O}_{td}.$ $\mathcal{B}$ responds $\mathcal{O}_{H}$ as in Theorem 1 and responds other oracles as follows. 

\begin{itemize}

	\item \textbf{Ciphertext Oracle} $\mathcal{O}_{ct}:$ Given a keyword set $\mathbb{W}=\{\omega_j\}_{j\in[m]}=\{n_j,v_j\}_{j \in[m]}$, $\mathcal{B}$ retrieves entry $(w_j,h_j,t_j,\delta_j)$ corresponding to each keyword in $\mathbb{W}$ from $L_H$. If there is an entry with $\delta_j=0$, $\mathcal{B}$ outputs a random guess for $\theta$ and aborts (event $\textsf{abort}_1$). Otherwise, $\mathcal{B}$ randomly picks $x_1,x_2,x_3 \in \mathbb{Z}_p^*$, sets 	
	$$ct_{1,j}=(g^s)^{x_3t_j}\cdot g^{x_1+x_2}, ct_2=(g^a)^{x_1},$$	
	$$ct_3=(pk_A)^{x_2}, ct_4=(g^s)^{x_3},$$
	and outputs $ct=(\{ct_{1,j}\}_{j\in [m]},ct_2,ct_3,ct_4)$.
	
	\item \textbf{Trapdoor Oracle} $\mathcal{O}_{td}:$ Given a keyword policy $\mathbb{P}=\{M,\pi,\{\pi(j)\}_{j\in[l]}\}$, $\mathcal{B}$ retrieves entry $(w_j,h_j,t_j,\delta_j)$ corresponding to each keyword in $\mathbb{P}$ from $L_H$. If there is an entry with $\delta_j=0$, $\mathcal{B}$ outputs a random guess for $\theta$ and aborts (event $\textsf{abort}_1$). Otherwise, $\mathcal{B}$ randomly selects $y_1,y_2,y_3\in \mathbb{Z}_p^*$ and $\textbf{v}\in \mathbb{Z}_p^{n-1}$, sets 
	
	$$td_{11,j}=g^{M_{j}(y_{1}+y_{2}\|\textbf{v})^{\mathrm{T}}}\cdot (g^b)^{y_3t_j}, td_{12,j}=(g^b)^{y_3t_j}\cdot g^{y_1+y_2},$$	
	$$td_2=(g^a)^{y_1}, td_3=(pk_A)^{y_2}, td_4=(g^b)^{y_3},$$
	and outputs $td=(\mathbb{P}=\{M,\pi,\{td_{12,j}\}_{j\in[l]}\},\{td_{11,j}\}_{j\in [l]},td_2,td_3,td_4)$.
\end{itemize}

\textbf{Challenge:} $\mathcal{A}$ selects two keyword policies of equal size, $\mathbb{P}_0^*=\{M_0^*,\pi_0^*,\{\pi_0^*(i)\}_{i\in[m]}\}$ and $\mathbb{P}_1^*=\{M_1^*,\pi_1^*,\{\pi_1^*(i)\}_{i\in[m]}\}$ with the restriction that neither of them is allowed to be queried in $\mathcal{O}_{td}$ or be satisfied by any keyword set that has been queried in $\mathcal{O}_{ct}$. It submits them to $\mathcal{B}$ as the challenge keyword policies. Next, $\mathcal{B}$ retrieves entry $(w_{ki},h_{ki},t_{ki},\delta_{ki})$ corresponding to each keyword in $\{\mathbb{P}_k^*\}_{k\in\{0,1\}}$ from $L_H$, where $i \in[m]$. We set $\{\Delta_k^*=null\}_{k\in\{0,1\}}$. If each $\delta_{ki}$ in the entry in $L_H$ corresponding to each $w_{ki}$(that's say, $\pi_k^*(i)$) in $\mathbb{P}_k^*$ is 0, we set $\Delta_k^*=0$; Otherwise, we set $\Delta_k^*=1$. Next, $\mathcal{B}$ works as follows.

(1) If $\Delta_0^*=\Delta_1^*=1$ or there exists $\Delta_k^*=null$, where $k\in\{0,1\}$, $\mathcal{B}$ aborts (event $\textsf{abort}_2$);

(2) Otherwise, there's a bit $\hat{\omega}$ such that $\Delta_{\hat{\omega}}^*=0$ with $h_{\hat{\omega}i}^*=g^{d/b}\cdot g^{t_{\hat{\omega}i}^*}$. Then $\mathcal{B}$ generates challenge trapdoor for $\mathbb{P}_{\hat{\omega}}^*$. It randomly selects $y_2^*,y_3^* \in \mathbb{Z}_p^*$, implicitly sets $y_1^*=cy_3^*$. It then selects $\textbf{v}_1,\textbf{v}_2\in \mathbb{Z}_p^{n-1}$,                                                                                                               implicitly makes $g^{M_{i}(dy_3^*\|\textbf{0})^{\mathrm{T}}}=g^{dy_3^*}$ hold. $\mathcal{B}$ next sets
$$td_{11,i}^*=\mathcal{Z}^{M_{i}(y_3^*\|\textbf{v}_1)^{\mathrm{T}}}\cdot g^{M_{i}(y_2^*\|\textbf{v}_2)^{\mathrm{T}}}\cdot(g^b)^{y_3^*t_{\hat{\omega}i}^*},$$
$$td_{12,i}^*=\mathcal{Z}^{y_3^*}\cdot(g^b)^{y_3^*t_{\hat{\omega}i}^*}\cdot g^{y_2^*}, td_2^*=(g^{ac})^{y_3^*},$$	
$$td_3^*=(pk_A)^{y_2^*}, td_4^*=(g^b)^{y_3^*},$$

and outputs $td^*=(\mathbb{P}_{\hat{\omega}}^*=\{M_{\hat{\omega}}^*,\pi_{\hat{\omega}}^*,\{td_{12,i}^*\}_{i\in[m]}\},\{td_{11,i}^*\}_{i\in [m]},td_2^*,td_3^*,td_4^*)$. 

If $\mathcal{Z}=g^{c+d}$, then $td^*$ is a well-formed trapdoor; else $\mathcal{Z}=g^z$ where $z$ is randomly chosen from $\mathbb{Z}_p^*$.

\textbf{Query phase \uppercase\expandafter{\romannumeral 2:} } $\mathcal{A}$ continues to issue queries with the same restriction in the challenge phase.

\textbf{Guess:} $\mathcal{A}$ returns the guess $\hat{\omega}' \in \{0,1\}$. If $\hat{\omega}'=\hat{\omega}$, $\mathcal{B}$ sets $\theta' =0$; otherwise, it set $\theta' =1$. 

Similar to Theorem 1, the probability that $\mathcal{B}$ succeeds in guessing the bit $\theta$ and solves the mDLIN problem is $ |\Pr[\theta'=\theta]-\frac{1}{2}|$, which is non-negligible.

\textit{Theorem 4:} The proposed scheme is TI-CS secure assuming the hardness of the mDLIN problem.

\textit{Proof:} The proof is similar to that of Theorem 1.

\textbf{Initialization:} $\mathcal{B}$ runs $\mathbf{Setup(1^\lambda)}\to \mathbb{PP}$, randomly chooses $s \in \mathbb{Z}_p^*$, sets $pk_S=g^s,sk_S=s,pk_A=g^a,pk_R=g^b,$ (which implies $sk_A=a,sk_R=b$). Then it sends $(\mathbb{PP},pk_A,pk_S,pk_R)$ to $\mathcal{A}$, who then submits $pk_C$ to $\mathcal{B}$.

\textbf{Query phase \uppercase\expandafter{\romannumeral 1:} } $\mathcal{A}$ submits polynomial queries to $\mathcal{O}_{H},\mathcal{O}_{ct},\mathcal{O}_{Tran_{ct}},\mathcal{O}_{td},\mathcal{O}_{Tran_{td}}.$ $\mathcal{B}$ responds $\mathcal{O}_{H}$ as in Theorem 1 and responds other oracles as follows. 

\begin{itemize}
	
	\item \textbf{Ciphertext Oracle} $\mathcal{O}_{ct}:$ Given a keyword set $\mathbb{W}=\{\omega_j\}_{j\in[m]}=\{n_j,v_j\}_{j \in[m]}$, $\mathcal{B}$ retrieves entry $(w_j,h_j,t_j,\delta_j)$ corresponding to each keyword in $\mathbb{W}$ from $L_H$. If there is an entry with $\delta_j=0$, $\mathcal{B}$ outputs a random guess for $\theta$ and aborts (event $\textsf{abort}_1$). Otherwise, $\mathcal{B}$ randomly picks $x_1,x_2,x_3 \in \mathbb{Z}_p^*$, sets 
	
	$$ct_{1,j}=(g^s)^{x_3t_j}\cdot g^{x_1+x_2}, ct_2=(pk_C)^{x_1},$$	
	$$ct_3=(g^a)^{x_2}, ct_4=(g^s)^{x_3},$$
	and $ct=(\{ct_{1,j}\}_{j\in [m]},ct_2,ct_3,ct_4)$, adds $(ct,\{t_j\}_{j\in[m]},x_1,x_2,x_3)$ in an initially empty list $L_{ct}$ and returns $ct$ to $\mathcal{A}$.
	
	\item \textbf{Transform Ciphertext Oracle} $\mathcal{O}_{Tran_{ct}}:$ Given a ciphertext $ct=(\{ct_{1,j}\}_{j\in [m]},ct_2,ct_3,ct_4)$, $\mathcal{B}$ retrieves entry $(ct,\{t_j\}_{j\in[m]},x_1,x_2,x_3)$ from $L_{ct}$. $\mathcal{B}$ randomly picks $x' \in \mathbb{Z}_p^*$ and implicitly sets $x'=\hat{x}a$, then sets
	
	$$ct_{1,j}'=ct_{1,j}^{x'}, ct_2'=ct_2^{x'},ct_3'=(g^{x_2})^{x'}, ct_4'=ct_4^{x'},$$
	and outputs $ct'=(\{ct_{1,j}'\}_{j\in [m]},ct_2',ct_3',ct_4')$.
	
	\item \textbf{Trapdoor Oracle} $\mathcal{O}_{td}:$ Given a keyword policy $\mathbb{P}=\{M,\pi,\{\pi(j)\}_{j\in[l]}\}$, $\mathcal{B}$ retrieves entry $(w_j,h_j,t_j,\delta_j)$ corresponding to each keyword in $\mathbb{P}$ from $L_H$. If there is an entry with $\delta_j=0$, $\mathcal{B}$ outputs a random guess for $\theta$ and aborts (event $\textsf{abort}_1$). Otherwise, $\mathcal{B}$ randomly selects $y_1,y_2,y_3\in \mathbb{Z}_p^*$ and $\textbf{v}\in \mathbb{Z}_p^{n-1}$, sets 
	
	$$td_{11,j}=g^{M_{j}(y_{1}+y_{2}\|\textbf{v})^{\mathrm{T}}}\cdot (g^b)^{y_3t_j}, td_{12,j}=(g^b)^{y_3t_j}\cdot g^{y_1+y_2},$$	
	$$td_2=pk_C^{y_1}, td_3=(g^a)^{y_2}, td_4=(g^b)^{y_3},$$
	
	and $td=(\mathbb{P}=\{M,\pi,\{td_{12,j}\}_{j\in[l]}\},\{td_{11,j}\}_{j\in [l]},td_2,td_3,td_4)$, adds $(td,\{t_j\}_{j\in[l]},y_1,y_2,y_3,\textbf{v})$  in an initially empty list $L_{td}$ and returns $td$ to $\mathcal{A}$.
	
	\item \textbf{Transform Trapdoor Oracle} $\mathcal{O}_{Tran_{ts}}:$ Given a trapdoor $td=(\mathbb{P}=\{M,\pi,\{td_{12,j}\}_{j\in[l]}\},\{td_{11,j}\}_{j\in [l]},td_2,td_3,td_4)$, $\mathcal{B}$ retrieves entry $(td,\{t_j\}_{j\in[l]},y_1,y_2,y_3,\textbf{v})$  from $L_{td}$. $\mathcal{B}$ randomly picks $y' \in \mathbb{Z}_p^*$ and implicitly sets $y'=\hat{y}a$, then sets
	
	$$td_{11,j}'=td_{11,j}^{y'}, td_{12,j}'=td_{12,j}^{y'},$$	
	$$td_2'=td_2^{y'}, td_3'=(g^{y_2})^{y'}, td_4'=td_4^{y'},$$
	
	and outputs $td'=(\mathbb{P}'=\{M,\pi,\{td_{12,j}'\}_{j\in[l]}\},\{td_{11,j}'\}_{j\in [l]},td_2',td_3',td_4')$.
	
\end{itemize}

\textbf{Challenge:} $\mathcal{A}$ selects two keyword policies of equal size, $\mathbb{P}_0^*=\{M_0^*,\pi_0^*,\{\pi_0^*(i)\}_{i\in[m]}\}$ and $\mathbb{P}_1^*=\{M_1^*,\pi_1^*,\{\pi_1^*(i)\}_{i\in[m]}\}$ with the restriction that $\mathcal{A}$ should not have queried any $(n_{\pi_0^*(i)},v_{\pi_0^*(i)})$ or $(n_{\pi_1^*(i)},v_{\pi_1^*(i)})$ in $\mathcal{O}_{ct}$ or $\mathcal{O}_{td}$. Next, $\mathcal{B}$ retrieves entry $(w_{ki},h_{ki},t_{ki},\delta_{ki})$ corresponding to each keyword in $\{\mathbb{P}_k^*\}_{k\in\{0,1\}}$ from $L_H$, where $i \in[m]$. We set $\{\Delta_k^*=null\}_{k\in\{0,1\}}$. If each $\delta_{ki}$ in the entry in $L_H$ corresponding to each $w_{ki}$ in $\mathbb{P}_k^*$ is 0, we set $\Delta_k^*=0$; Otherwise, we set $\Delta_k^*=1$. Next, $\mathcal{B}$ works as follows.

(1) If $\Delta_0^*=\Delta_1^*=1$ or there exists $\Delta_k^*=null$, where $k\in\{0,1\}$, $\mathcal{B}$ aborts (event $\textsf{abort}_2$);

(2) Otherwise, there's a bit $\hat{\omega}$ such that $\Delta_{\hat{\omega}}^*=0$ with $h_{\hat{\omega}i}^*=g^{d/b}\cdot g^{t_{\hat{\omega}i}^*}$. Then $\mathcal{B}$ generates challenge trapdoor for $\mathbb{P}_{\hat{\omega}}^*$. It randomly selects $y_1^*,y_3^* \in \mathbb{Z}_p^*$, implicitly sets $y_2^*=cy_3^*$. It then selects $\textbf{v}_1,\textbf{v}_2\in \mathbb{Z}_p^{n-1}$,                                                                                                               implicitly makes $g^{M_{i}(dy_3^*\|\textbf{0})^{\mathrm{T}}}=g^{dy_3^*}$ hold. $\mathcal{B}$ next sets
$$td_{11,i}^*=\mathcal{Z}^{M_{i}(y_3^*\|\textbf{v}_1)^{\mathrm{T}}}\cdot g^{M_{i}(y_1^*\|\textbf{v}_2)^{\mathrm{T}}}\cdot(g^b)^{y_3^*t_{\hat{\omega}i}^*},$$
$$td_{12,i}^*=\mathcal{Z}^{y_3^*}\cdot(g^b)^{y_3^*t_{\hat{\omega}i}^*}\cdot g^{y_1^*}, td_2^*=(pk_C)^{y_1^*},$$	
$$td_3^*=(g^{ac})^{y_3^*}, td_4^*=(g^b)^{y_3^*},$$

and outputs $td^*=(\mathbb{P}_{\hat{\omega}}^*=\{M_{\hat{\omega}}^*,\pi_{\hat{\omega}}^*,\{td_{12,i}^*\}_{i\in[m]}\},\{td_{11,i}^*\}_{i\in [m]},td_2^*,td_3^*,td_4^*)$. 

If $\mathcal{Z}=g^{c+d}$, then $td^*$ is a well-formed trapdoor; else $\mathcal{Z}=g^z$ where $z$ is randomly chosen from $\mathbb{Z}_p^*$.

\textbf{Query phase \uppercase\expandafter{\romannumeral 2:} } $\mathcal{A}$ continues to issue queries with the same restriction in the challenge phase.

\textbf{Guess:} $\mathcal{A}$ returns the guess $\hat{\omega}' \in \{0,1\}$. If $\hat{\omega}'=\hat{\omega}$, $\mathcal{B}$ sets $\theta' =0$; otherwise, it set $\theta' =1$. 

Similar to Theorem 1, the probability that $\mathcal{B}$ succeeds in guessing the bit $\theta$ and solves the mDLIN problem is $ |\Pr[\theta'=\theta]-\frac{1}{2}|$, which is non-negligible.

\section{Further Discussion} \label{Further Discussion}

In our system, the transformed ciphertext can be utilized to construct an inverted index without decryption, similar to \cite{DBLP:journals/tifs/LiHHS23}, to improve the search efficiency of the cloud server. In this section, we define a new ciphertext structure to perform an equality test on each encrypted keyword, which we call it single transformed ciphertext $ct^*_i$. Specifically, a single transformed ciphertext contains $(ct_{1,i}',ct_2',ct_3',ct_4')$, corresponding to each keyword in $\mathbb{W}=\{\omega_i\}_{i\in[m]}$. Accordingly, we define a new trapdoor structure to perform an equality test on each encrypted keyword, which we call it single transformed trapdoor $td^*_i$. Specifically, a single transformed ciphertext contains $(td_{12,i}',td_2',td_3',td_4')$, corresponding to each keyword in $\mathbb{P}=\{M,\pi,\{td_{12,i}\}_{i\in[l]}\}$. The process of generating an inverted index when the cloud server is offline and searching using the inverted index when online is as follows:

\begin{algorithm}[H]
	\caption{InsertIndex}\label{alg:alg1}
	\begin{algorithmic}
		\Require $\mathbb{PP}, ct', \mathcal{H} = <\text{Label, Pointer}>$
		\Ensure $\mathcal{H}$
		
		\For {each $ct^*_i$ in $ct'$} 
		\State ptr $\gets \mathcal{H}$;
		\State isEqual $\gets 0$;
		\While {ptr $\neq$ Null}
		\State isEqual $\gets$ \textbf{EqualTest}($\mathbb{PP}$, $ct^*_i$, ptr.Label, $sk_C$);
		\If {isEqual $\neq$ 0} 
		\State insert $ct'$ into ptr.Pointer;  
		\State break;
		\EndIf
		\State ptr $\gets$ ptr.NextNode();
		\EndWhile
		
		\If {isEqual == 0} 
		\State create newNode;
		\State newNode.Label $\gets$ $ct^*_i$;
		\State insert $ct'$ into ptr.Pointer;
		\State insert newNode at the end of $\mathcal{H}$;
		\EndIf
		\EndFor
		
		\State \Return $\mathcal{H}$
	\end{algorithmic}
	\label{alg1}
\end{algorithm}

\begin{algorithm}[H]
	\caption{FastSearch}\label{alg:alg1}
	\begin{algorithmic}
		\Require $\mathcal{PP}, td', \mathcal{H} = <\text{Label, Pointer}>$
		\Ensure $\mathcal{R}$
		\State $\mathcal{C} \gets \emptyset$;
		\For{each $td_i^*$ in $td'$} 
		\State ptr $\gets$ $\mathcal{H}$
		\While{ptr $\neq$ Null}
		\If{\textbf{MatchTest}($\mathbb{PP},ptr.Label,td_i^*,sk_C)$} 
		\State tempPtr $\gets$ ptr.Pointer 
		\While{tempPtr $\neq$ Null} 
		\State $\mathcal{C} \gets \mathcal{C} \cup \{tempPtr.FilePointer\}$ 
		\State tempPtr $\gets$ tempPtr.$P_Next$
		\EndWhile
		\State break; 
		\EndIf
		\State ptr $\gets$ ptr.NextNode();
		\EndWhile
		\EndFor
		\State $\mathcal{R} \gets \emptyset$;
		\For{each $ct'$ in $\mathcal{C}$}
		\If{$\mathbf{Search}(sk_{C},ct',td')$ $==$ True}
		\State $\mathcal{R} \gets \mathcal{R} \cup \{ct'\}$ 
		\EndIf
		\EndFor
		
		\State \Return $\mathcal{R}$
	\end{algorithmic}
	\label{alg1}
\end{algorithm}

\begin{itemize}
	\item  $\mathbf{EqualTest}( \mathbb{PP},ct^{*1}_i,ct^{*2}_j,sk_C)$: Given $\mathbb{PP}$, two single transformed ciphertext, parse $ct^{*1}_i,ct^{*2}_j$ as $(ct_{1,i}'^{1},ct_2'^{1},ct_3'^{1},ct_4'^{1})$ and $(ct_{1,j}'^{2},ct_2'^{2},ct_3'^{2},ct_4'^{2})$, in which $ct_{1,i}'^{1}$ and $ct_{1,j}'^{2}$ are single terms from the encrypted keyword set $\{ct_{1,i}'^{1}\}_{i\in [m_1]}$ and $\{ct_{1,j}'^{2}\}_{j\in [m_2]}$, respectively. By running this algorithm, the cloud server uses its private key $sk_C$ to determine whether the two single transformed ciphertext contain the same keyword. It return 1 if the following equation holds, and 0 otherwise.
	
	$$e(\frac{ct_{1,i}^{\prime1_{sk_C}}}{ct_{2}^{\prime1}ct_{3}^{\prime1_{sk_C}}},ct_4'^{2})=e(\frac{ct_{1,j}^{\prime2_{sk_C}}}{ct_{2}^{\prime2}ct_{3}^{\prime2_{sk_C}}},ct_4'^{1})$$
\end{itemize}	

\begin{itemize}
	\item  $\mathbf{MatchTest}( \mathbb{PP},ct^{*1}_i,td^{*2}_j,sk_C)$: Given $\mathbb{PP}$, a single transformed ciphertext and a single transformed trapdoor, parse $ct^{*1}_i,td^{*2}_j$ as $(ct_{1,i}'^{1},ct_2'^{1},ct_3'^{1},ct_4'^{1})$ and $(td_{12,j}'^{2},td_2'^{2},td_3'^{2},td_4'^{2})$, in which $ct_{1,i}'^{1}$ and $td_{12,j}'^{2}$ are single terms from the encrypted keyword set $\{ct_{1,i}'^{1}\}_{i\in [m]}$ and $\{td_{12,j}'^{2}\}_{j\in [l]}$, respectively. By running this algorithm, the cloud server uses its private key $sk_C$ to determine whether the single transformed ciphertext and the single transformed trapdoor contain the same keyword. It return 1 if the following equation holds, and 0 otherwise.	
	$$e(\frac{ct_{1,i}^{\prime1_{sk_C}}}{ct_{2}^{\prime1}ct_{3}^{\prime1_{sk_C}}},td_{4}^{\prime2})=e(\frac{td_{12,j}^{\prime2 sk_C}}{td_{2}^{\prime2}td_{3}^{\prime2 sk_C}},ct_{4}^{\prime1})$$	
	\item $\mathbf{InitIndex}( \mathbb{PP})$: Given $\mathbb{PP}$, this algorithm initializes an empty inverted index $\mathcal{I}$ with header $\mathcal{H}=<\text{Label, Pointer}>$.

	\item $\mathbf{InsertIndex}( \mathbb{PP},ct',\mathcal{H})$: Given $\mathbb{PP}$, a transformed ciphertext $ct'$, the index $\mathcal{I}$, this algorithm inserts $ct'$ into $\mathcal{I}$ as in Algorithm 1 and after several rounds of insertion, $\mathcal{I}$ can be instanced as in Fig 3.
	
	\item $\mathbf{FastSearch}( \mathbb{PP},td',\mathcal{H})$: Given $\mathbb{PP}$, a transformed trapdoor $td'$, the index $\mathcal{H}$, this algorithm returns a collection of matched documents $\mathcal{R}$, each of which satisfies the search query of the receiver, as shown in Algorithm 2.
	
\end{itemize}

\begin{figure}[htbp]
	\centering
	\includegraphics[width=0.5\textwidth]{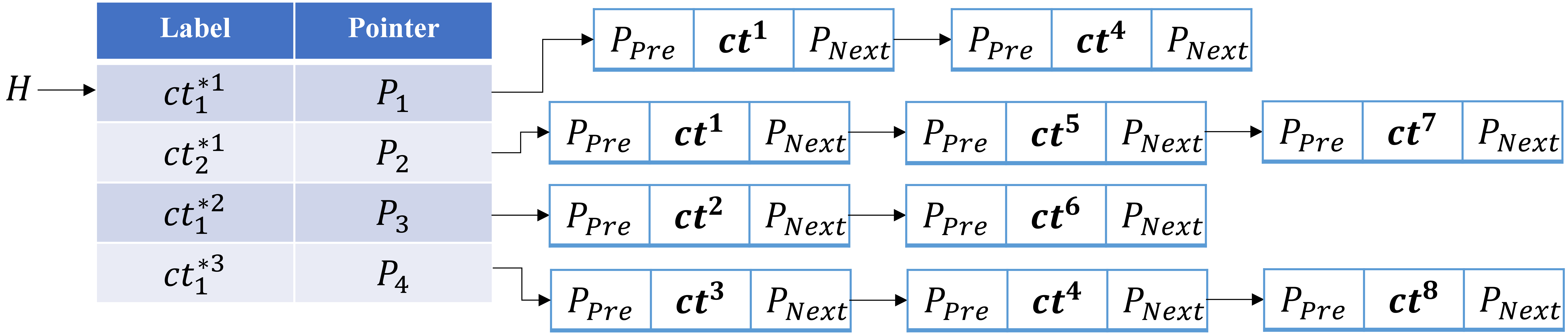}
	\caption{Inverted index.\label{fig:3}}
\end{figure}

\section{Performance} \label{Performance}

In this section, we compare the communicational and computational overheads of our scheme with other existing schemes and experimentally evaluate our scheme.

\subsection{Theoretical analysis}
We compare the properties with the related state-of-the-art works\cite{DBLP:journals/tdsc/CuiWDWL18}\cite{DBLP:journals/istr/TsengFL22}\cite{DBLP:journals/jsa/HuangHLHL23}\cite{DBLP:journals/tifs/LiHHS23}	\cite{DBLP:journals/tifs/ChengM24}\cite{DBLP:conf/uss/Meng0TML24}, as shown in Table \ref{tab2}.
Let $|\mathbb{G}|$ and $|\mathbb{G}_T|$ be the size of the elements in the groups $\mathbb{G}$ and $\mathbb{G}_T$, respectively, $m$ be the size of the set of keywords belonging to the ciphertext, $|\mathbb{P}|$ be the size of the access matrix, $l$ be the number of keywords in the access structure, and $t$ be the number of columns in the access matrix. For ease of comparison, in the scheme that does not support expressive queries, we use $n$ to represent the number of multi-keywords in its query, which is of the same order of magnitude as $l$. Denote $\mathcal{P}$ as a pairing operation, $\mathcal{E}$ as an exponentiation operation, $\mathcal{M}$ as a multiplication operation, $\mathcal{D}$ as a division operation, $\mathcal{H}$ as a hash to point operation, $X_1$ as the number of elements in $I_{\mathbb{M},\rho} = \{\mathcal{I}_1, \ldots, \mathcal{I}_{X_1}\}$ and $X_2$ as $|\mathcal{I}_1| + \cdots + |\mathcal{I}_{X_1}|$.

\begin{table*}[!t]
	\caption{Property comparison\label{tab2}}
	\renewcommand{\arraystretch}{1}
	\setlength{\tabcolsep}{4pt}
	\centering
	\begin{tabular}{|c|c|c|c|c|c|c|c|}
		\hline
		
		Schemes & Expressiveness & KGA & \makecell[c]{Ciphertext Efficiency \\When Multiple Receivers} &\makecell[c]{Trapdoor Efficiency\\When Multiple Senders}  &\makecell[c]{Search \\Efficiency}  &\makecell[c]{Keyword Name \\Privacy Protection}  & \makecell[c]{Do Not Need Trusted \\Trapdoor Generator}\\
		\hline
		CWDWL16\cite{DBLP:journals/tdsc/CuiWDWL18}  & AND,OR & Partial & \ding{51} &\ding{51} &\ding{55}
		&\ding{55} &\ding{55}   \\ 
		\hline
		TFL22\cite{DBLP:journals/istr/TsengFL22}  & AND,OR & \ding{55} & \ding{55} & \ding{51} &\ding{55}
		&\ding{55} &\ding{51}    \\ 
		\hline
		HHLHL23\cite{DBLP:journals/jsa/HuangHLHL23}  & AND & \ding{51} & \ding{55} & \ding{55} &\ding{51}&\ding{51} &\ding{51}  \\ 
		\hline
		LHHS23\cite{DBLP:journals/tifs/LiHHS23}& AND & \ding{51} & \ding{51} & \ding{51} & \ding{51} &\ding{51}&\ding{55}  \\ 
		\hline
		CM24\cite{DBLP:journals/tifs/ChengM24} & AND & \ding{51} & \ding{51} & \ding{51} & \ding{51} &\ding{51}&\ding{51} \\ 
		\hline
		MCTML24\cite{DBLP:conf/uss/Meng0TML24} & AND,OR & \ding{51} & \ding{55} & \ding{55} & \ding{55} & \ding{55}&\ding{51} \\
		\hline
		Ours & AND,OR,Threshold & \ding{51} & \ding{51} & \ding{51} &\ding{51} &\ding{51} &\ding{51}   \\
		
		\hline
	\end{tabular}
\end{table*}

Table \ref{tab3} provides a theoretical evaluation of the communication overhead for a single sender and a single receiver in the system, where the sender generates a set of ciphertexts corresponding to a set of keywords for an EHR, and the receiver generates expressive/and queries for multiple keywords. It can be seen that our scheme has a smaller ciphertext overhead than \cite{DBLP:journals/istr/TsengFL22} and \cite{DBLP:conf/uss/Meng0TML24} encryption of keyword sets compared to other schemes. When performing multi-keyword queries of the same order of magnitude ($n$ and $l$), we have a smaller overhead gap with several schemes\cite{DBLP:journals/jsa/HuangHLHL23}\cite{DBLP:journals/tifs/LiHHS23}\cite{DBLP:journals/tifs/ChengM24}\cite{DBLP:conf/uss/Meng0TML24}. The reason is that compared with\cite{DBLP:journals/jsa/HuangHLHL23}\cite{DBLP:journals/tifs/LiHHS23} \cite{DBLP:journals/tifs/ChengM24}, we support expressive queries, and compared with\cite{DBLP:conf/uss/Meng0TML24}, we can provide privacy protection for keyword names.

Table \ref{tab4} shows the communication overhead when there are \textbf{S} senders and \textbf{R} receivers, that is, in a large-scale user system. Due to the introduction of a fully trusted trapdoor generation server, the communication overhead of \cite{DBLP:journals/tdsc/CuiWDWL18} is independent of the number of users, but this may cause system bottlenecks. \cite{DBLP:journals/istr/TsengFL22}'s invariant trapdoor overhead is determined by its inability to resist KGA; that is, the receiver only needs to use its own private key to encrypt the query, but the number of its keyword ciphertexts is linearly related to the number of receivers in the system. The communication overheads of \cite{DBLP:journals/jsa/HuangHLHL23} and \cite{DBLP:conf/uss/Meng0TML24} are both linearly related to the number of users in the system. Only \cite{DBLP:journals/tifs/LiHHS23} and \cite{DBLP:journals/tifs/ChengM24} can provide ciphertext and trapdoor overheads that are independent of the number of users in the system. However, they cannot provide expressive queries, which highlights the advantage of our scheme.

Table \ref{tab5} provides a theoretical evaluation of the computational overhead of several major algorithms for a single sender and a single receiver in the system. The computational overhead of the sender in our scheme running the encryption algorithm on the keyword set is comparable to \cite{DBLP:conf/uss/Meng0TML24} and lower than other schemes. At the same time, the computational overhead of the receiver generating trapdoors is comparable to \cite{DBLP:journals/jsa/HuangHLHL23} and \cite{DBLP:journals/tifs/LiHHS23} and lower than other schemes. In the testing phase, due to the support of expressive queries, the computational overhead of our scheme and \cite{DBLP:journals/tdsc/CuiWDWL18}\cite{DBLP:journals/istr/TsengFL22}\cite{DBLP:conf/uss/Meng0TML24} are higher than other schemes. In addition, due to the support of access policies and privacy protection of keyword names in ciphertext, our computational overhead is relatively within a reasonable range.

We further analyze the computational overhead of users in the case of multiple senders and receivers in the system in Table \ref{tab6}. Just like the communication overhead, the computational overhead of the sender in \cite{DBLP:journals/istr/TsengFL22} is linearly related to the number of receivers. In addition, the computational overhead of \cite{DBLP:journals/jsa/HuangHLHL23} and \cite{DBLP:conf/uss/Meng0TML24} schemes also increases linearly as the number of users in the system gradually increases. The computational overhead of user encryption in our scheme is better than \cite{DBLP:journals/tifs/LiHHS23} \cite{DBLP:journals/tifs/ChengM24}, and the trapdoor generation overhead is similar to these two.

\begin{table}[h]
	\caption{Communication Cost For Single Sender and Single Receiver\label{tab3}}
	\renewcommand{\arraystretch}{0.5}
	\setlength{\tabcolsep}{11pt}
	\begin{tabularx}{\columnwidth}{ccc}
		\toprule
		\textbf{Schemes} & 	\textbf{Ciphertext}  & 	\textbf{Trapdoor}  \\
		\midrule
		CWDWL16\cite{DBLP:journals/tdsc/CuiWDWL18}  & $(5m+1)|\mathbb{G}|+|\mathbb{G}_T|$  & $(6l+2)|\mathbb{G}|+|\mathbb{P}|$   \\ 
		\midrule
		TFL22\cite{DBLP:journals/istr/TsengFL22}  & $(m+1)|\mathbb{G}|+2|\mathbb{G}_T|$ & $(l^2+l)|\mathbb{G}|+|\mathbb{P}|$  \\ 
		\midrule
		HHLHL23\cite{DBLP:journals/jsa/HuangHLHL23} & $2m|\mathbb{G}|$ & $2n|\mathbb{G}|$  \\ 
		\midrule
		LHHS23\cite{DBLP:journals/tifs/LiHHS23} & $5m|\mathbb{G}|$ & $2n|\mathbb{G}|$  \\ 
		\midrule
		CM24\cite{DBLP:journals/tifs/ChengM24}  & $3m|\mathbb{G}|$ & $3n|\mathbb{G}|$ \\ 
		\midrule
		MCTML24\cite{DBLP:conf/uss/Meng0TML24}  & $(m+2)|\mathbb{G}|+|\mathbb{G}_T|$ & $(2l+1)|\mathbb{G}|+|\mathbb{P}|$  \\
		\midrule
		Ours & $2|\mathbb{G}|$  & $(2l+3)|\mathbb{G}|+|\mathbb{P}|$  \\
		\bottomrule
	\end{tabularx}
\end{table}

\begin{table}[h]
	\caption{Communication Cost For \textbf{S} Sender and \textbf{R} Receiver\label{tab4}}
	\renewcommand{\arraystretch}{0.5}
	\setlength{\tabcolsep}{9pt}
	\begin{tabularx}{\columnwidth}{ccc}
		\toprule
		\textbf{Schemes} & 		\textbf{Ciphertext}  & 	\textbf{Trapdoor}  \\
		\midrule
		CWDWL16\cite{DBLP:journals/tdsc/CuiWDWL18}  &  $(5m+1)|\mathbb{G}|+|\mathbb{G}_T|$  & $(6l+2)|\mathbb{G}|+|\mathbb{P}|$   \\ 
		\midrule
		TFL22\cite{DBLP:journals/istr/TsengFL22}  & $(m+1)\textbf{R}|\mathbb{G}|+2\textbf{R}|\mathbb{G}_T|$ & $(l^2+l)|\mathbb{G}|+|\mathbb{P}|$  \\ 
		\midrule
		HHLHL23\cite{DBLP:journals/jsa/HuangHLHL23}  & $2m\textbf{R}|\mathbb{G}|$ & $2n\textbf{S}|\mathbb{G}|$  \\ 
		\midrule
		LHHS23\cite{DBLP:journals/tifs/LiHHS23}& $5m|\mathbb{G}|$ & $2n|\mathbb{G}|$  \\ 
		\midrule
		CM24\cite{DBLP:journals/tifs/ChengM24} & $3m|\mathbb{G}|$ & $3n|\mathbb{G}|$ \\ 
		\midrule
		MCTML24\cite{DBLP:conf/uss/Meng0TML24} & $(m+2)\textbf{R}|\mathbb{G}|+\textbf{R}|\mathbb{G}_T|$ & $(2l+1)\textbf{S}|\mathbb{G}|+|\mathbb{P}|$  \\
		\midrule
		Ours & $(m+3)|\mathbb{G}|$ & $(2l+3)|\mathbb{G}|+|\mathbb{P}|$  \\
		\bottomrule
	\end{tabularx}
\end{table}

\begin{table*}[h]
	\caption{Computation Cost For Single Sender and Single Receiver\label{tab5}}
	\renewcommand{\arraystretch}{0.5}
	\setlength{\tabcolsep}{16.5pt}
	\begin{tabularx}{\textwidth}{cccc}
		\toprule
		\textbf{Schemes} & 		\textbf{Encryption}  & 	\textbf{Trapdoor Generation}  & 	\textbf{Test}  \\
		\midrule
		CWDWL16\cite{DBLP:journals/tdsc/CuiWDWL18}  &  $(6m+3)\mathcal{E}+2m\mathcal{M}$  & \makecell[c]{$1\mathcal{P}+(3+10l)\mathcal{E}$ \\ $+5l\mathcal{M}+1\mathcal{H}$}   & \makecell[c]{$\leq (6X_2+1)\mathcal{P}+(X_2+1)\mathcal{E}$ \\ $+5X_2\mathcal{M}+X_2\mathcal{D}+1\mathcal{H}$}    \\ 
		\midrule
		TFL22\cite{DBLP:journals/istr/TsengFL22}  &\makecell[c]{$(2m+3)\mathcal{E}+(m+1)\mathcal{M}$ \\ $+m\mathcal{H}$}   & $(5l-1)\mathcal{E}+2l\mathcal{M}+l\mathcal{H}$ & \makecell[c]{$\leq 2X_1\mathcal{P}+2X_2\mathcal{E}$ \\ $+(2X_2+X_1)\mathcal{M}+2X_1\mathcal{D}$}    \\ 
		\midrule
		HHLHL23\cite{DBLP:journals/jsa/HuangHLHL23}   & $(2m+1)\mathcal{E}+m\mathcal{H}$ & $(2n+1)\mathcal{E}+n\mathcal{D}+n\mathcal{H}$ & $2n\mathcal{P}$    \\  
		\midrule
		LHHS23\cite{DBLP:journals/tifs/LiHHS23}& $9m\mathcal{E}+2m\mathcal{M}+5m\mathcal{H}$ & $2n\mathcal{E}+n\mathcal{M}+n\mathcal{H}$  & $2n\mathcal{P}$   \\ 
		\midrule
		CM24\cite{DBLP:journals/tifs/ChengM24} &$4m\mathcal{E}+m\mathcal{H}$ & $4n\mathcal{E}+n\mathcal{H}$  & $4n\mathcal{P}+2n\mathcal{E}+2n\mathcal{D}$    \\ 
		\midrule
		MCTML24\cite{DBLP:conf/uss/Meng0TML24} &$(m+3)\mathcal{E}+1\mathcal{D}+m\mathcal{H}$   &\makecell[c]{$(4l+1)\mathcal{E}+(tl+l+1)\mathcal{M}$ \\ $+2\mathcal{D}+l\mathcal{H}$}   & \makecell[c]{$\leq 3X_1\mathcal{P}+3X_2\mathcal{E}$\\
			$+(X_1+3X_2)\mathcal{M}+X_1\mathcal{D}$}
		\\
		\midrule
		Ours &\makecell[c]{$(m+4)\mathcal{E}+(m+1)\mathcal{M}$ \\ $+m\mathcal{H}$}  &\makecell[c]{$(2l+4)\mathcal{E}+(tl+2l+1)\mathcal{M}$ \\ $+l\mathcal{H}$}  &\makecell[c]{$\leq (m+l+2X_1)\mathcal{P}+(m+l+3X_2+2)\mathcal{E}$ \\ $+(2+3X_2)\mathcal{M}+(m+l+X_1+X_2)\mathcal{D}$}      \\
		\bottomrule
	\end{tabularx}
\end{table*}

\begin{table}[h]
	\caption{Computation Cost For \textbf{S} Sender and \textbf{R} Receiver\label{tab6}}
	\renewcommand{\arraystretch}{0.5}
	\setlength{\tabcolsep}{1pt}
	\begin{tabularx}{\columnwidth}{ccc}
		\toprule
		\textbf{Schemes} & 		\textbf{Encryption}  & 	\textbf{Trapdoor Generation}  \\
		\midrule
		CWDWL16\cite{DBLP:journals/tdsc/CuiWDWL18}  &  $(6m+3)\mathcal{E}+2m\mathcal{M}$  & \makecell[c]{$1\mathcal{P}+(3+10l)\mathcal{E}$ \\ $+5l\mathcal{M}+1\mathcal{H}$} \\ 
		\midrule
		TFL22\cite{DBLP:journals/istr/TsengFL22}  &\makecell[c]{$(2m+3)\textbf{R}\mathcal{E}$ \\ $+(m+1)\textbf{R}\mathcal{M}+m\textbf{R}\mathcal{H}$}   & $(5l-1)\mathcal{E}+2l\mathcal{M}+l\mathcal{H}$ \\ 
		\midrule
		HHLHL23\cite{DBLP:journals/jsa/HuangHLHL23}  & $(2m+1)\textbf{R}\mathcal{E}+m\textbf{R}\mathcal{H}$ & $(2n+1)\textbf{S}\mathcal{E}+n\textbf{S}\mathcal{D}+n\mathcal{H}$ \\ 
		\midrule
		LHHS23\cite{DBLP:journals/tifs/LiHHS23}& $9m\mathcal{E}+2m\mathcal{M}+5m\mathcal{H}$ & $2n\mathcal{E}+n\mathcal{M}+n\mathcal{H}$   \\
		\midrule
		CM24\cite{DBLP:journals/tifs/ChengM24} &$4m\mathcal{E}+m\mathcal{H}$ & $4n\mathcal{E}+n\mathcal{H}$  \\ 
		\midrule
		MCTML24\cite{DBLP:conf/uss/Meng0TML24} &\makecell[c]{$(m+3)\textbf{R}\mathcal{E}+\textbf{R}\mathcal{D}$\\ $+m\textbf{R}\mathcal{H}$} &\makecell[c]{$(4l+1)\textbf{S}\mathcal{E}+(tl+l+1)\textbf{S}\mathcal{M}$ \\ $+2\textbf{S}\mathcal{D}+l\textbf{S}\mathcal{H}$}  \\
		\midrule
		Ours &\makecell[c]{$(m+4)\mathcal{E}+(m+1)\mathcal{M}$ \\ $+m\mathcal{H}$}  &\makecell[c]{$(2l+4)\mathcal{E}$ \\ $+(tl+2l+1)\mathcal{M}+l\mathcal{H}$} \\
		\bottomrule
	\end{tabularx}
\end{table}

\begin{figure}[htbp]
	\centering
	\includegraphics[width=0.5\textwidth]{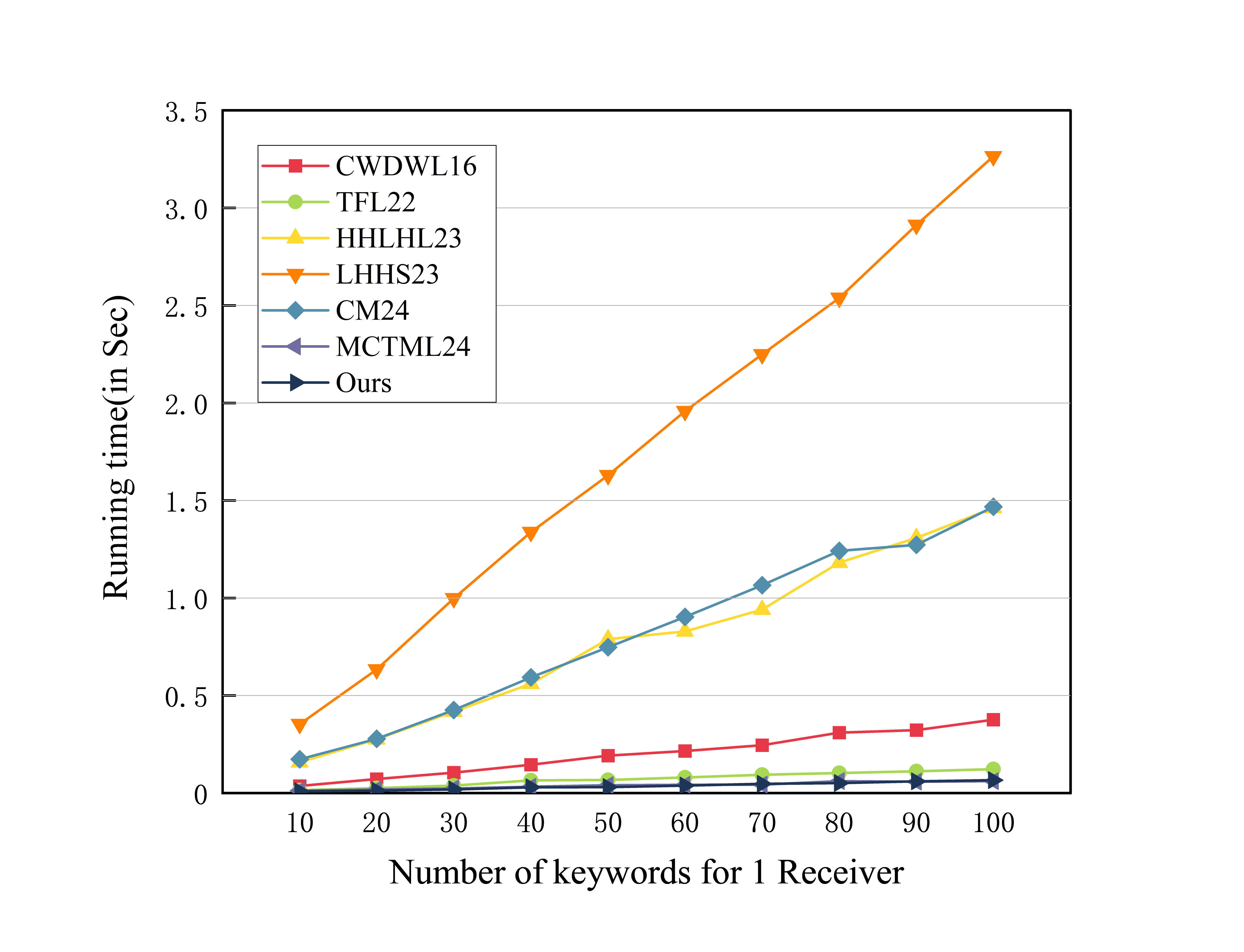}
	\caption{Encryption time for a sender with only one receiver.\label{fig:3}}
\end{figure}

\subsection{Experimental Results}

We developed a prototype of the encryption scheme using the MNT224 curve in the Charm 0.5 \cite{DBLP:conf/ndss/Akinyele0R12} framework using Python 3.10.12. All running times involved in the experiments were measured on a PC equipped with an Intel (R) CoreTM i3-12100 CPU @ 3.30 GHz and 16.0 GB of RAM. Since all Charm routines were designed under the asymmetric group, our construction was converted to an asymmetric setting before implementation, similar to \cite{DBLP:journals/tdsc/CuiWDWL18}.

Figure \ref{fig:3} shows the encryption time for each sender in different schemes when there is only one receiver in the system. The time overhead of all schemes increases linearly with the size of the keyword set. It can be seen that our scheme has the smallest time overhead among all schemes. Since only single-keyword AND gate search is supported, the time overhead and encryption cost of \cite{DBLP:journals/jsa/HuangHLHL23},\cite{DBLP:journals/tifs/LiHHS23} and \cite{DBLP:journals/tifs/ChengM24}increase faster as the keyword set grows, because a separate ciphertext needs to be generated for each keyword.

\begin{figure}[htbp]
	\centering
	\includegraphics[width=0.5\textwidth]{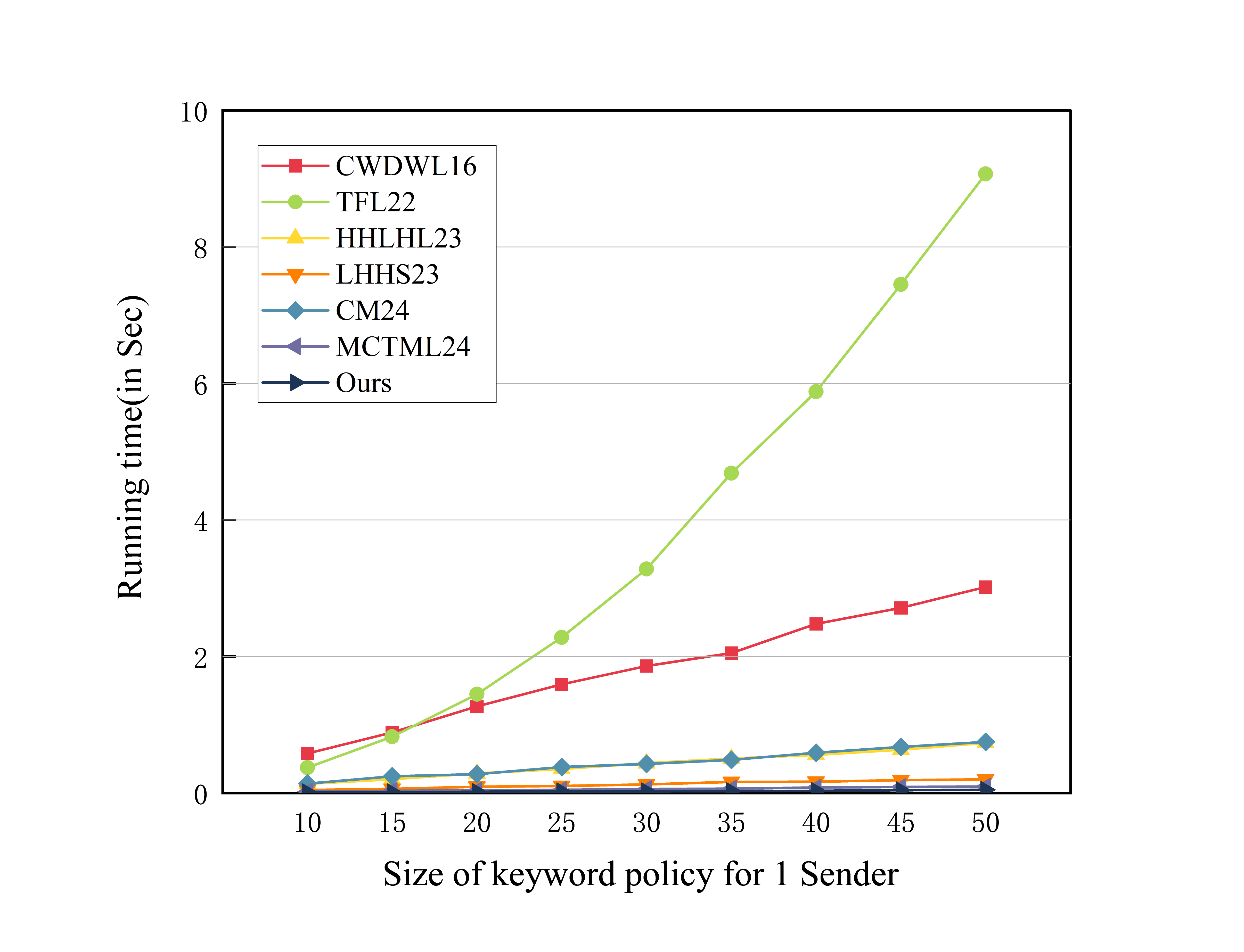}
	\caption{Trapdoor generation time for a receiver with only one sender.\label{fig:4}}
\end{figure}

Figure \ref{fig:4} shows the encryption time for each receiver in different schemes when there is only one sender in the system. The time overhead of all schemes increases linearly with the size of the keyword strategy. It can be seen that our scheme has the smallest time overhead among all schemes. Corresponding to the algorithm theoretical analysis in Table \ref{tab5}, the trapdoor generation time overhead of schemes \cite{DBLP:journals/tdsc/CuiWDWL18} and \cite{DBLP:journals/istr/TsengFL22} is larger. Since encryption is required for each keyword in the policy, the overhead in schemes \cite{DBLP:journals/jsa/HuangHLHL23},\cite{DBLP:journals/tifs/LiHHS23} and \cite{DBLP:journals/tifs/ChengM24} are higher than that of \cite{DBLP:conf/uss/Meng0TML24} and ours.

\begin{figure}[htbp]
	\centering
	\includegraphics[width=0.5\textwidth]{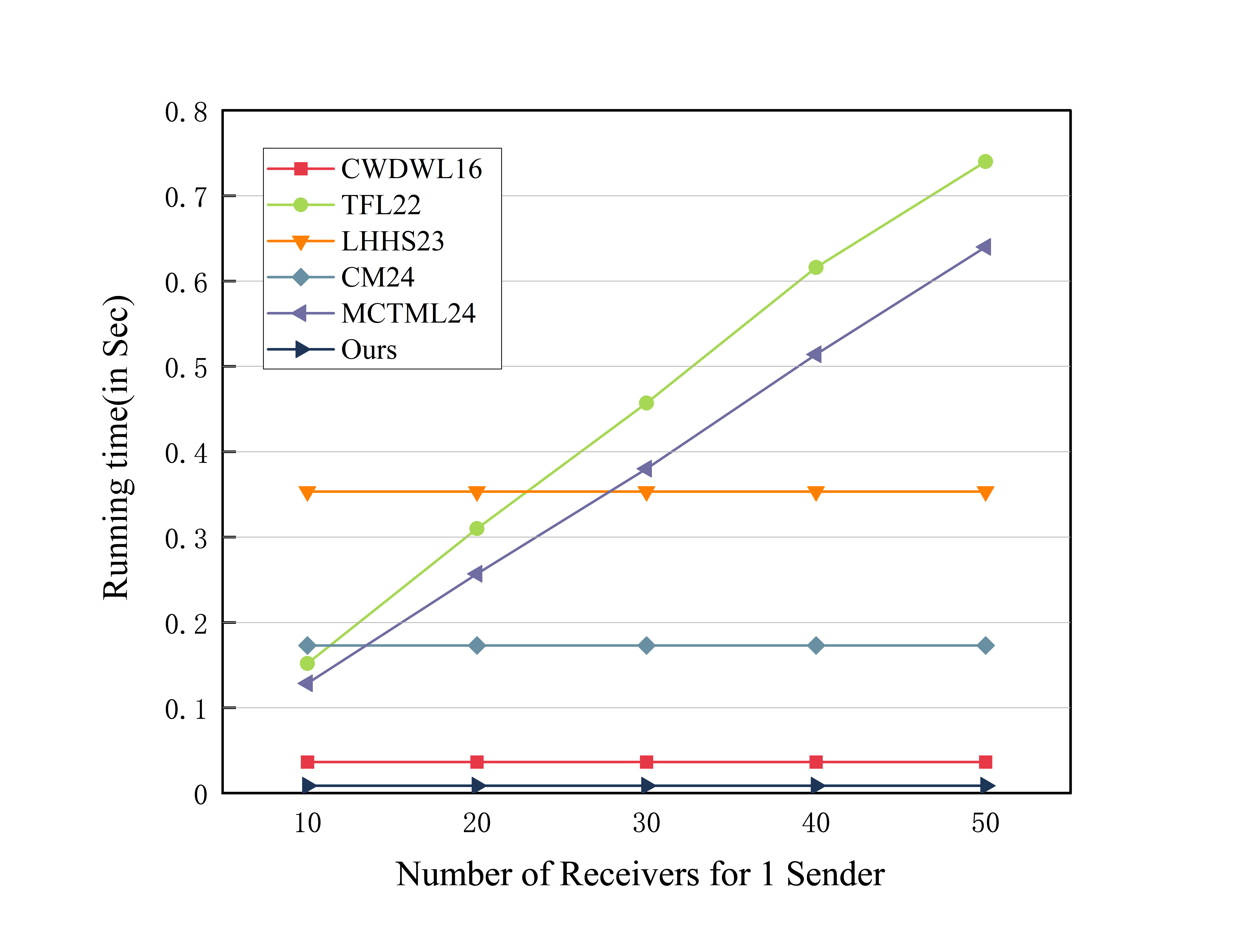}
	\caption{Encryption time for a sender with multiple receivers(assumes that the keyword set size is fixed to 10).\label{fig:5}}
\end{figure}

\begin{figure}[htbp]
	\centering
	\includegraphics[width=0.5\textwidth]{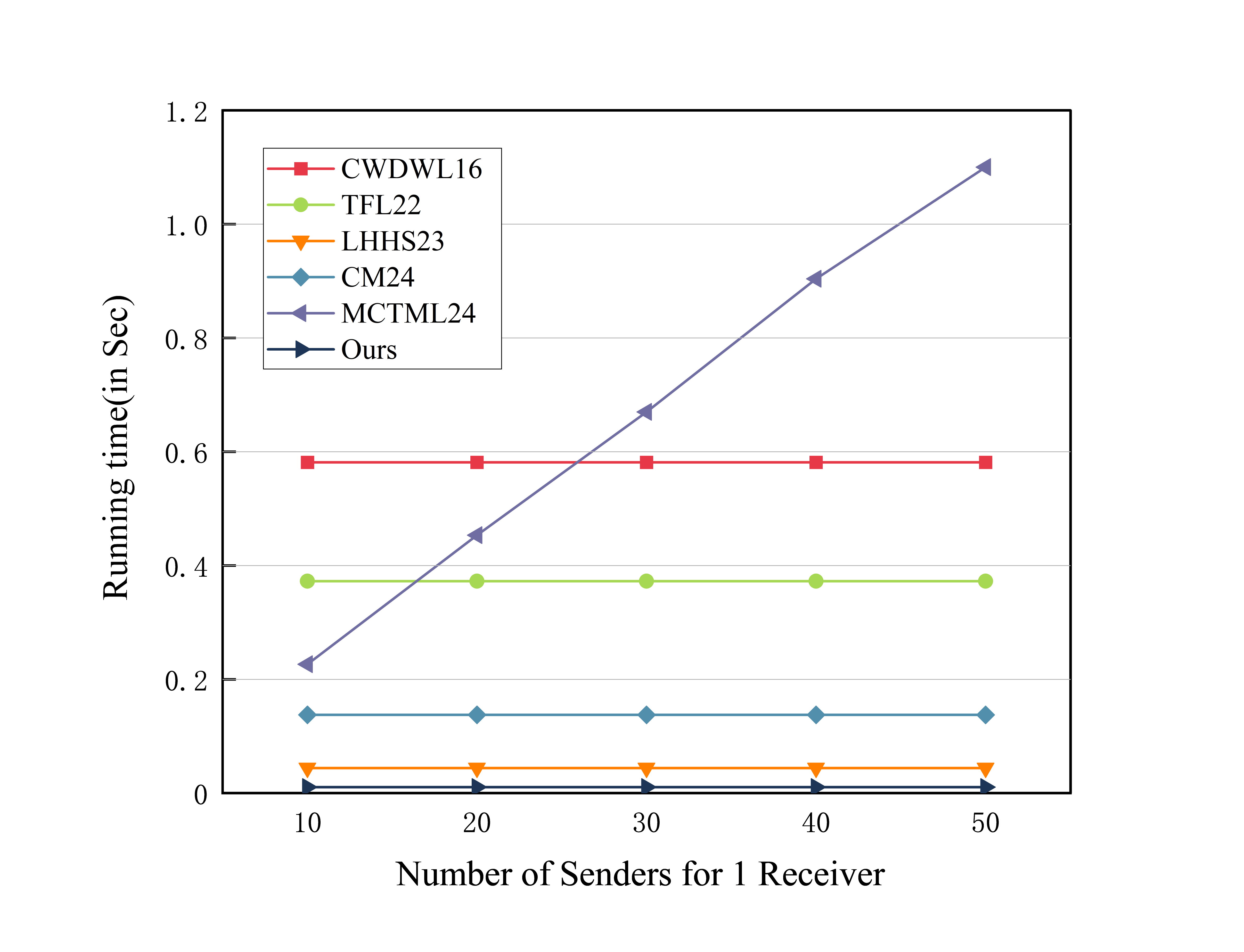}
	\caption{Trapdoor generation time for a receiver with multiple senders(assumes that the keyword policy size is fixed to 10).\label{fig:6}}
\end{figure}
Figure \ref{fig:5} shows the encryption time for each sender in different schemes when there are multiple receivers in the system, assuming that the keyword set size is constant at 10. Since the encryption algorithm supports encryption independent of the number of users in the system, the time overhead of \cite{DBLP:journals/tdsc/CuiWDWL18}, \cite{DBLP:journals/tifs/LiHHS23}, \cite{DBLP:journals/tifs/ChengM24} and our scheme is independent of the number of receivers, and our scheme has the smallest time overhead. Corresponding to the algorithm theoretical analysis in Table \ref{tab6}, the encryption time overhead of \cite{DBLP:journals/istr/TsengFL22} and \cite{DBLP:conf/uss/Meng0TML24} both grow linearly with the number of system receivers. Since \cite{DBLP:journals/jsa/HuangHLHL23} only supports single keyword and single user search, its overhead for multiple keywords and multiple receivers is much greater than that of other schemes.

Figure \ref{fig:6} shows the encryption time for each receiver in different schemes when there are multiple senders in the system, relative to Figure \ref{fig:5}, assuming that the keyword policy size is constant at 10. Since the encryption algorithm that supports encryption is independent of the number of users in the system, the time overhead of \cite{DBLP:journals/tdsc/CuiWDWL18}, \cite{DBLP:journals/tifs/LiHHS23},\cite{DBLP:journals/tifs/ChengM24},\cite{DBLP:journals/istr/TsengFL22} and our scheme is independent of the number of receivers, and our scheme has the smallest time overhead. Among them, since \cite{DBLP:journals/istr/TsengFL22} scheme cannot resist KGA attack, the sender's public key is not needed when generating trapdoors, so the trapdoor overhead is also independent of the sender. Corresponding to the theoretical analysis of the algorithm in Table \ref{tab6}, the encryption time overhead of \cite{DBLP:conf/uss/Meng0TML24} grows linearly with the number of system and receivers.

\section{Conclusion}

We propose an efficient and expressive PAEKS scheme for scenarios with a large number of users, such as EHR. Its innovation lies in its ability to simultaneously achieve low user overhead when sharing searchable data while supporting expressive query capabilities. We only introduced one auxiliary server, which can reduce user overhead and achieve efficient queries while resisting KGA. In addition, unlike the previous partial hiding, our scheme can also achieve complete privacy protection for plaintext keyword names under expressive queries. We conducted security analysis and experimental performance analysis on the system scheme, demonstrating the excellent performance of the system. Currently, most PEKS schemes only support unilateral access control. In the future, we will expand our research work to PEKS that can achieve bilateral access control, realizing a more flexible, user-friendly searchable encryption system.

\section*{Acknowledgments}
This should be a simple paragraph before the References to thank those individuals and institutions who have supported your work on this article.


\bibliographystyle{IEEEtran}
\bibliography{reference}  

\newpage

\vfill

\end{document}